\documentclass[acmsmall,screen,nonacm]{acmart}

\usepackage{tikz}
\usepackage[inline]{enumitem}
\usepackage{booktabs}
\usepackage{amsmath}
\usepackage{relsize}
\usepackage[capitalise, nameinlink]{cleveref}

\Crefname{figure}{Figure}{Figures}
\Crefname{section}{Section}{Sections}
\usepackage{multicol}
\usepackage{multirow}
\usepackage{units}
\usepackage{numprint}
\usepackage{subfigure}
\usepackage{soul}
\usepackage{array}
\usepackage{algorithm}
\usepackage[noend]{algpseudocode}
\usepackage{listings}
\usepackage{svg}
\usepackage{balance}
\usepackage{wrapfig}

\usetikzlibrary{
    matrix,
    decorations.pathreplacing, 
    calc, 
    positioning,
    fit, 
    backgrounds, 
    shapes, 
    shapes.misc,
    patterns}

\definecolor{BerlinBlack}{RGB}{0,0,6}
\definecolor{BerlinOrange}{RGB}{255, 108,0}
\definecolor{BerlinWhite}{RGB}{255,255,255}
\definecolor{BerlinGray}{RGB}{110, 110, 114}
\definecolor{BerlinRed}{RGB}{196, 13, 30}
\definecolor{BifoldBlue}{RGB}{0,45,96}
\definecolor{BifoldLightBlue}{RGB}{42,186,212}
\definecolor{BifoldLightLightBlue}{RGB}{127,204,224}
\definecolor{BifoldLightestBlue}{RGB}{177,222,235}

\definecolor{blue}{RGB}{42,186,212}
\definecolor{red}{RGB}{196, 13, 30}
\definecolor{green}{HTML}{2ca02c}
\definecolor{orange}{HTML}{ff7f0e}
\definecolor{purple}{HTML}{9467bd}
\definecolor{brown}{HTML}{8c564b}
\definecolor{pink}{HTML}{e377c2}
\definecolor{gray}{HTML}{7f7f7f}
\definecolor{olive}{HTML}{bcbd22}
\definecolor{cyan}{HTML}{17becf}

\definecolor{mplBlue}{HTML}{1f77b4}
\definecolor{mplOrange}{HTML}{ff7f0e}
\definecolor{mplGreen}{HTML}{2ca02c}
\definecolor{mplRed}{HTML}{d62728}
\definecolor{mplPurple}{HTML}{9467bd}
\definecolor{mplBrown}{HTML}{8c564b}
\definecolor{mplPink}{HTML}{e377c2}
\definecolor{mplGray}{HTML}{7f7f7f}
\definecolor{mplOlive}{HTML}{bcbd22}
\definecolor{mplCyan}{HTML}{17becf}

\definecolor{commentgreen}{RGB}{106,168,79}

\newcommand{\Qv}{\mathcal{Q}}

\newcommand{\dv}{\Delta}

\newcommand{\mat}[1]{\ensuremath{\mathbf{#1}}}

\npstyleenglish 

\newcommand{\eat}[1]{}

\newcommand{\redd}[1]{}

\ExplSyntaxOn
\cs_new:Npn \expandableinput #1
{ \use:c { @@input } { \file_full_name:n {#1} } }
\AddToHook{env/tabular/begin}
{ \cs_set_eq:NN \input \expandableinput }
\ExplSyntaxOff

\definecolor{codegreen}{rgb}{0,0.6,0}
\definecolor{codegray}{rgb}{0.5,0.5,0.5}
\definecolor{codepurple}{rgb}{0.58,0,0.82}
\definecolor{backcolour}{rgb}{0.95,0.95,0.92}

\lstdefinestyle{mystyle}{
    basicstyle=\ttfamily\tiny,
    breakatwhitespace=false,
    breaklines=true,
    keepspaces=true,
	lineskip=-.04cm,
    numbersep=4pt,
    showspaces=false,
    showstringspaces=false,
    showtabs=false,
    tabsize=2
}

\tikzset{roundnode/.style={circle, draw=orange!60, 
    fill=orange!10, thick, minimum size=2.5mm, inner sep=0.0, font=\rmfamily\fontsize{7}{7}\selectfont}}

\newcommand{\nodeLabel}[1]{\begin{tikzpicture}
        \node[roundnode,  inner sep = 0, outer sep=0]{#1};
    \end{tikzpicture}}

\tikzset{roundnodeb/.style={roundnode, draw=brown!80, fill=brown!10}}

\newcommand{\nodeLabelb}[1]{\begin{tikzpicture}
    \node[roundnodeb,  inner sep = 0, outer sep=0]{#1};
\end{tikzpicture}}

\newcommand{\CodeSymbol}[1]{\bfseries\textcolor{green}{#1}}   

\lstdefinestyle{highlightStyle}{
    basicstyle=\ttfamily\fontsize{10pt}{10pt},
    breakatwhitespace=false,
    breaklines=false,
    keepspaces=true,
    lineskip=-.04cm,
    numbersep=4pt,
    showspaces=false,
    showstringspaces=false,
    showtabs=false,
    tabsize=2,
    otherkeywords={read, transformencode, augment, print, lmCG},
    morekeywords={read, transformencode, augment, print, lmCG},
    keywordstyle=\color{brown}\textbf,
    classoffset=2, 
    otherkeywords={parfor},
    morekeywords={parfor},
    keywordstyle=\color{purple}\textbf,
    frame = lines,
    literate={\{}{{\CodeSymbol{\{}}}1
           {\}}{{\CodeSymbol{\}}}}1
           {(}{{\CodeSymbol{(}}}1
           {)}{{\CodeSymbol{)}}}1
}

\tikzset{ts/.style={text height = 0.5ex, text depth=.08ex}}

\tikzset{cg/.style={
            matrix of nodes, anchor=north,
            every node/.style={inner sep=0.5mm}
        }}
\tikzset{cgs/.style={
            matrix of nodes, anchor=north,
            every node/.style={anchor=base, ts},
            nodes={}
        }}
\tikzset{mlabel/.style={yshift=0.1cm}}
\tikzset{dictL/.style={right, align=right, xshift= 1.cm, node distance=1cm, ts}}
\tikzset{dict/.style={
            ts,
            right,
            xshift= -1cm,
            text width=2.4ex,  node distance=1cm}}
\tikzset{bbCus/.style={draw, inner sep=-0.5mm, rounded corners}}

\tikzset{bbt1/.style={
            draw,
            inner sep=0.3mm,
            rounded corners,
            fill = white,
        }}

\tikzset{bbt2/.style={
    draw,
    inner sep=0.6mm,
    rounded corners,
    fill = white,
}}

\tikzset{
    hyperlink node/.style={
        alias=sourcenode,
        append after command={
            let     \p1 = (sourcenode.north west),
                \p2=(sourcenode.south east),
                \n1={\x2-\x1},
                \n2={\y1-\y2} in
            node [inner sep=0pt, outer sep=0pt,anchor=north west,at=(\p1)] {\hyperlink{#1}{\XeTeXLinkBox{\phantom{\rule{\n1}{\n2}}}}}
        }
    }
}

\tikzset{tiles/.style={
            matrix of nodes, anchor=north,
            every node/.style={fulln},
            row sep=0.1em, column sep=0.1em
        }}

\tikzset{fulln/.style={ minimum height=2em,
						minimum width=2em,
						fill=blue!50, draw,
						inner sep=0.5mm,
						rounded corners}}

\tikzset{tiles/.style={
						matrix of nodes, anchor=north,
						every node/.style={fulln},
						row sep=0.1em, column sep=0.1em
				}}
\tikzset{clusterNode/.style={ inner sep=0.5mm,
						minimum height=2em,
						minimum width=6em,
						fill=black!10, draw,
						inner sep=0.5mm,
						rounded corners}}

\tikzset{cluster/.style={
						matrix of nodes, anchor=north,
						every node/.style={clusterNode},
						row sep=0.1em, column sep=0.1em
				}}

\tikzset{infoBox/.style={
                        rounded corners,
                        minimum height = 2em,
                        minimum width = 1.7cm,
                        draw }}

\tikzset{frame/.style={
            matrix of nodes, anchor=north,
            every node/.style={inner sep=0.5mm, minimum height=0.9em},
        }}

\tikzset{fontscale/.style = {font=\relsize{#1}}}

\tikzset{Mtiles/.style={
						matrix of nodes, anchor=north,
						every node/.style={fulln},
						row sep=0.1em, column sep=0.1em
				}}

\tikzset{infoBox/.style={
						rounded corners,
						minimum height = 2em,
						minimum width = 1.7cm,
						draw
				}}

\tikzset{pointer/.style={
                    fulln, 
                    minimum height=0.2cm,
                    minimum width =0.2cm,
                    inner sep=0.25mm,
                    rounded corners=2.8pt
                }}

\newcommand{\tikzarrowfig}{\begin{tikzpicture}
	\draw[->, thick](-0.15,-0.02) -- (0.15,-0.02);
	\node[](0,0.15){ };
\end{tikzpicture}}

\tikzset{fulln/.style={ inner sep=0.5mm,
						minimum height=1.4em,
						minimum width=1.4em,
						fill=blue!50, draw,
						inner sep=0.5mm,
						rounded corners = 1mm}}

\tikzset{tiles/.style={
						matrix of nodes, anchor=north,
						every node/.style={fulln},
						row sep=0.1em, column sep=0.1em,
				}}
\tikzset{clusterNode/.style={ inner sep=0.5mm,
						minimum height=1.6em,
						minimum width=3.5em,
						fill=black!10, draw,
						inner sep=0.5mm,
						rounded corners = 1mm}}

\tikzset{cluster/.style={
						matrix of nodes, anchor=north,
						every node/.style={clusterNode},
						row sep=0.1em, column sep=0.1em
				}}

\tikzset{smalln/.style={ fulln}}

\tikzset{smalltiles/.style={tiles,
				every node/.style={smalln},
		}}

\begin{document}

\title{Morphing-based Compression for Data-centric ML Pipelines}

\author{Sebastian Baunsgaard}
\orcid{0009-0001-1463-7294}
\affiliation{
  \institution{Technische Universität Berlin}
  \city{Berlin}
  \country{Germany}
}

\author{Matthias Boehm}
\orcid{0000-0003-1344-3663}
\affiliation{
  \institution{Technische Universität Berlin}
  \city{Berlin}
  \country{Germany}
}

\renewcommand{\shortauthors}{Baunsgaard and Boehm}

\begin{abstract}
Data-centric ML pipelines extend traditional machine learning (ML) pipelines---of feature transformations and ML model training---by outer loops for data cleaning, augmentation, and feature engineering to create high-quality input data. Existing lossless matrix compression applies lightweight compression schemes to numeric matrices and performs linear algebra operations such as matrix-vector multiplications directly on the compressed representation but struggles to efficiently rediscover structural data redundancy. Compressed operations are effective at fitting data in available memory, reducing I/O across the storage-memory-cache hierarchy, and improving instruction parallelism.
The applied data cleaning, augmentation, and feature transformations provide a rich source of information about data characteristics such as distinct items, column sparsity, and column correlations. 
In this paper, we introduce BWARE---an extension of AWARE for workload-aware lossless matrix compression---that pushes compression through feature transformations and engineering to leverage information about structural transformations. Besides compressed feature transformations, we introduce a novel technique for lightweight morphing of a compressed representation into workload-optimized compressed representations without decompression. 
BWARE shows substantial end-to-end runtime improvements, reducing the execution time for training data-centric ML pipelines from days to hours.
\end{abstract}

\maketitle

\vspace{-0.15cm}
\section{Introduction}
\label{sec:introduction}

Modern machine learning (ML) training comprises more than just selecting and fitting ML algorithms or neural network architectures as well as their hyper-parameters. Data-centric ML pipelines extend traditional ML pipelines of feature transformations and model training by additional pre-processing steps for data validation \cite{SchelterLSCBG18, GrafbergerGS23}, data cleaning \cite{SiddiqiKB23}, feature engineering \cite{SalazarNA21}, and data augmentation \cite{RatnerBEFWR2017, RatnerDWSR2016, VarmaR18, KrizhevskySH12} to construct high-quality datasets with good coverage of the target domain. These pre-processing techniques can substantially improve model accuracy \cite{KrizhevskySH12, SiddiqiKB23} as well as other measures like fairness \cite{SalazarNA21, StoyanovichAHJS22} and robustness \cite{TsiprasSETM19}.

\textbf{Sources of Redundancy:} The iterative nature of finding good data-centric ML pipelines causes both operational redundancy (e.g., fully or partially repeated pre-processing steps) \cite{PhaniRB21} as well as data redundancy \cite{BaunsgaardB23}. Besides natural data redundancy, such as the small cardinality of categorical features and column correlations, data-centric ML pipelines create additional redundancy. Examples are the construction of new data points or features, as well as systematic transformations such as the imputation of missing values by mean or mode and data cleaning by robust functional dependencies \cite{DengFAWSEIMO017}. While being beneficial for model quality, the iterative selection of such data-centric ML pipelines is a very expensive process. Eliminating unnecessary redundancy through data reorganization is appealing because reorganization overheads can be amortized.   

\textbf{Lossless Matrix Compression:} A common approach for exploiting data redundancy without quality degradation is lossless compression. First, sparsity exploitation avoids processing zero values via dedicated data layouts, sparse operators, and even sparsity-exploiting ML algorithms \cite{YangWL20}. Common layouts include compressed sparse rows (CSR), columns (CSC), or coordinate format (COO)~\cite{IntelMKL, CuSparse, Saad90, SommerBERH19, KhamisNNOS20}. Second, existing compression techniques apply lightweight database compression schemes---such as dictionary encoding, run-length encoding, and offset-list encoding---to numeric matrices and perform linear algebra (LA) operations such as matrix multiplications directly on the compressed representation. Example frameworks are Compressed Linear Algebra (CLA)~\cite{ElgoharyBHRR18, ElgoharyBHRR16}, Tuple-oriented Compression (TOC)~\cite{FenganLYANPM19}, Grammar-compressed Matrices (GCM)~\cite{FerraginaMGTKNST22}, and AWARE~\cite{BaunsgaardB23} (see \cref{fig:Contribution}, top). AWARE creates compressed matrices ($\mathbb{C}$) in a workload-aware manner by (1) extracting a workload summary of the program at compile-time, as well as (2) workload-aware compression and compression-aware recompilation at runtime. Existing work struggles though to efficiently rediscover structural data redundancy in data-centric ML.

\setlength{\columnsep}{0.4cm}%
\begin{wrapfigure}{r}{8.cm}
		\centering
		\begin{tikzpicture}[ampersand replacement=\&]

				\node[bbt1](IO){\input{fig/prim/disk.tex}};

				\node[bbt1, xshift=1.05cm, yshift = 1cm, ](M){\input{fig/prim/Matrix.tex}};
				\node[bbt1, xshift=1.05cm, yshift = -1cm, ](F){\begin{tikzpicture}[ampersand replacement=\&]
    \node{
        \resizebox{1cm}{1cm}
        {
        \begin{tikzpicture}[ampersand replacement=\&]
            \matrix[Mtiles,
                column  1/.style={every node/.style={fulln, minimum width=1em, minimum height=4em, fill=green!50}},
                column  2/.style={every node/.style={fulln, minimum width=1em, minimum height=4em, fill=yellow!50}},
                column  3/.style={every node/.style={fulln, minimum width=1em, minimum height=4em, fill=red!50}},
             ](TILED){
                \  \& \  \& \ \\
            };
        \end{tikzpicture}
        }
    };
    \node at(0,0.7){Frame};
\end{tikzpicture}};

				\node[bbt2, xshift=0.5cm, fill=red!30, anchor = west](TRANSFORM){Transform};

				\draw[thick,->] ([yshift=-0.2cm]F.north east)  to[out = 0, in = -90]([xshift=0.6cm]TRANSFORM.south);
				\draw[thick,<-] ([yshift=0.2cm]M.south east)  to[out = 0, in = 90]([xshift=0.6cm]TRANSFORM.north);
				\draw[thick, <->] (IO) to[out = 090, in = 180] (M);
				\draw[thick, <->] (IO) to[out=-90, in=180] (F);

				\begin{scope}[yshift=1cm, xshift = 3cm] 
					
					\node[bbt1, xshift=1.6cm](CM){
\begin{tikzpicture}[ampersand replacement=\&]
    \node{
        \resizebox{1cm}{1cm}
        {
            \begin{tikzpicture}[ampersand replacement=\&]
                \matrix[Mtiles,
                column  1/.style={every node/.style={fulln, minimum width=1em, minimum height=1em, fill=blue!50}},
                column  2/.style={every node/.style={fulln, minimum width=1em, minimum height=4em, fill=blue!50}},
                column  3/.style={every node/.style={fulln, minimum width=1em, minimum height=2em, fill=blue!50}},
             ](TILED){
                \  \& \  \& \ \\
            };
            \end{tikzpicture}
        }
    };
    \node at(0,0.7){$\mathbb{C}$Matrix};
\end{tikzpicture}
};
					
					\node[bbt1, xshift=3.4cm](AL){\input{fig/prim/algo.tex}};

					\node[bbt2, yshift = -0.5cm](COMP){$\mathbb{C}$};
					\draw[thick, ->] (M) to[out = 0, in = 180] (COMP);
					\draw[thick, ->] (COMP) to[out = 0, in = 180] (CM);
					\draw[thick, ->] (CM) -- (AL);
					\node[bbt2, yshift =  0.1cm](WORKLOAD){Workload};
					\node[ xshift = -0.7cm, yshift=0.5cm, anchor=west](AWARE){\fontsize{9pt}{9pt}\textbf{AWARE}};
			
					\draw[thick, ->, dotted] (WORKLOAD) to[out = -90, in = 90] (COMP);
				\end{scope}

				\begin{scope}[yshift=-1cm, xshift=2.93cm]
					\node[bbt2, xshift=-.5cm,yshift = 0.9cm, fill=green!30, anchor = west](TRANSFORMC){$\mathbb{C}$Transform};
					\node[xshift = -0.7cm, yshift=-0.45cm,anchor=west](BWARE){\fontsize{9pt}{9pt}\textbf{BWARE}};
					\node[bbt1, xshift=1.6cm](CF){\begin{tikzpicture}[ampersand replacement=\&]
    \node{
        \resizebox{1cm}{1cm}
        {
        \begin{tikzpicture}[ampersand replacement=\&]
            \matrix[Mtiles,
                column  1/.style={every node/.style={fulln, minimum width=1em, minimum height=1em, fill=green!50}},
                column  2/.style={every node/.style={fulln, minimum width=1em, minimum height=4em, fill=yellow!50}},
                column  3/.style={every node/.style={fulln, minimum width=1em, minimum height=2em, fill=red!50}},
             ](TILED){
                \  \& \  \& \ \\
            };
        \end{tikzpicture}
        }
    };
    \node at(0,0.7){$\mathbb{C}$Frame};
\end{tikzpicture}};
					\draw[thick, ->] (F) -- (CF);
					\node[bbt2, xshift=0cm](COMP2){$\mathbb{C}$};

					\draw[thick, ->] ([yshift = 0.3cm] F.east) to[out =   0, in = -90] ([xshift= -0.3cm]TRANSFORMC.south);
					\draw[thick, ->] ([yshift = 0.3cm]CF.west) to[out = 180, in = -90] ([xshift= 0.3cm]TRANSFORMC.south);
					\draw[thick, ->] (TRANSFORMC.east) to[out = 0, in = -90] ([xshift= -0.1cm]CM.south);
					
					\node[bbt1, xshift=2.7cm, yshift = -0.3cm](IOC){\input{fig/prim/disk.tex}};
					\draw[thick, <->, ] ([xshift=0.1cm]IOC.north) 
					to[out = 90, in = -20] ([xshift=0.cm, yshift = 0.15cm]CF.north east)
					to[out = 160, in = -90] ([xshift=0.15cm]CM.south);
					\draw[thick, <->, ] ([xshift=-0.1cm]IOC.north) 
					to[out = 90, in = 0] ([yshift=0.3cm]CF.east);

					\node[bbt1, xshift=3.8cm,  ](MO){\input{fig/prim/morph.tex}};
					\draw[thick, <->, ] ([xshift= -0.1cm]MO.north) to[out = 90, in = -90] ([xshift=0.4cm]CM.south);

					\draw[thick, ->, dotted]([xshift = -0.2cm]WORKLOAD.north east) 
						to[out = 90, in = 180] ([yshift = 0.1cm]CM.north west)
						to[out = 0, in = 180] ([yshift = 0.1cm]AL.north east)
						to[out = 0, in = 90] ([xshift = 0.1cm]AL.north east)
						to[out = -90, in = 90] ([xshift = 0.1cm]AL.south east)
						to[out = -90, in = 90] ([xshift = 0.1cm]MO.north);

				\end{scope}

				\begin{scope}[on background layer]
					\node[bbCus, thin,fit=(BWARE) (TRANSFORMC) (MO), fill=blue!10,inner sep=1.5, rounded corners=1mm](BWAREBB){};
					\node[bbCus, thin,fit=(CM) (AWARE) (COMP) (WORKLOAD), fill=black!10,inner sep=1.5, rounded corners=1mm](AWAREBB){};
				\end{scope}

		\end{tikzpicture}
		\vspace{-0.3cm}
		\caption{\label{fig:Contribution}BWARE Framework Overview and Contributions. }
		\Description{...}
		
\end{wrapfigure}

\textbf{A Case for Compressed Pre-processing:} Feature transformations encode categorical and numerical features into numerical matrices.
This conversion is a rich source of information about structural data redundancy. For example, one-hot encoding a categorical feature requires determining the dictionary of $d$ distinct items and creating $d$ perfectly correlated binary features. Transformations like binning and feature hashing represent user-defined, lossy decisions which give upper bounds for code word sizes as well. Furthermore, data-centric ML pipelines iteratively evaluate additional features and different transformations. Therefore, we make a case for \emph{pushing compression through feature transformations and feature engineering to the sources}, e.g., storage. Holistic support requires (1) compressing the input frames in a form amenable to compressed feature transformations, (2) supporting compressed I/O, as well as (3) compressed feature engineering and feature transformations. Since data and workload characteristics of enumerated ML pipelines may differ, there is a need for morphing~\cite{HabichDUPKHL19, DammeUAJADL20, DammeDW15} compressed intermediate matrices into workload-optimized representations~\cite{BaunsgaardB23}.

\textbf{Contributions:} In this paper, we introduce BWARE (see \cref{fig:Contribution}, bottom) as a holistic, lossless compression framework for data-centric ML pipelines. Our main technical contributions are:
\begin{itemize}
	\item A lightweight frame compression scheme with dictionary encodings, enabling compressed feature transformations on heterogeneous data (\cref{sec:frameCompression}).
	\item A morphing technique for workload-aware tuning of compressed representations (\cref{sec:morphing}).
	\item Parallel and distributed I/O for compressed blocks without decompression (\cref{sec:io}).
	\item An optimizing compiler, injecting morphing instructions into LA programs (\cref{sec:compiler}).
	\item An experimental evaluation that studies the impact of compressed I/O, feature engineering, feature transformations, and training in data-centric ML pipelines (\cref{sec:experiments}).
\end{itemize}

\section{Potential Analysis}
\label{sec:potentialAnalysis}

We aim to quantify the potential of exploiting structural and value-based data redundancy.
To this end, we first summarize data characteristics of real-world datasets and investigate the potential runtime impact of pushing compression through pre-processing primitives.

\setlength{\columnsep}{0.6cm}

\subsection{Distinct Values}

\begin{figure}[b]
  	\centering
	\includegraphics[scale=0.99]{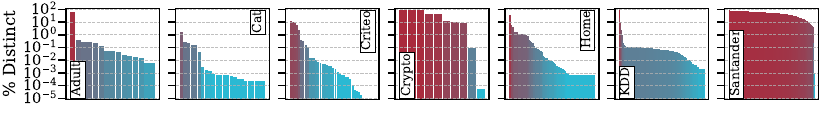}
	\vspace{-0.6cm}
	\caption{\label{fig:RelativeDistinct}Relative Number of Distinct Values in ML Datasets. Columns Sorted by the Number of Distinct Values.}
	\Description{...}
	\vspace{-0.25cm}
\end{figure}
The number of distinct values $d$ is a classic statistic to exploit in compression \cite{DammeDJW17, MullerCF14}, especially in string columns. Dense Dictionary Compression (DDC) constructs a dictionary of $d$ values and encodes the values as integers referencing positions in the dictionary. 
\cref{fig:RelativeDistinct} shows several datasets with their ratio of number of distinct values to rows per column. Some columns contain less than $0.001\%$ distinct values. Compressed operations that exploit the distinct values can reduce execution time in such cases by $99.999\%$~\cite{BaunsgaardB23}. Unfortunately, data does not always contain a low number of unique values, motivating additional compression techniques.

\begin{wrapfigure}{r}{6.5cm}
	\vspace{-0.5cm}
	\centering 
	\includegraphics[scale=0.99]{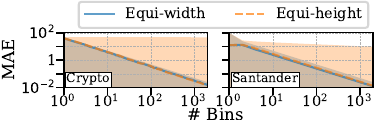}
	\vspace{-0.1cm}
	\caption{Lossy Quantization Effect on Values.}
	\label{fig:lossyBinning}
	\Description{...}
\end{wrapfigure}

\subsection{Lossy Transformations}

Feature engineering can reduce the number of distinct vales $d$ via lossy transformations such as binning, feature hashing, or quantization. We distinguish static and learned~\cite{ZhangLKADLZ17, ZhangYYH18} quantization schemes. An example of a static scheme is equi-width quantization, which scales the input values to discretized bins in the range of min-max with $\Qv_\dv(\mat{X}) = \hat{\mat{X}} = \lfloor \dv (\mat{X} - \mat{X}_{min}) / (\mat{X}_{max} - \mat{X}_{min})\rfloor$. The resulting number of distinct values is $d\leq\dv$, where $\dv$ is the configured number of bins. Increasing $\dv$ generally improves the accuracy of the approximation of the original data. $\dv=256$ is a common configuration, which allows encoding values in \texttt{UINT8}. In contrast, learned schemes~\cite{ZhangYYH18} use various techniques---such as quantiles or neural networks---to find optimal quantization boundaries (smaller bins for high-frequency value ranges). Equi-height quantization maps input values to buckets by $\dv$ quantiles. \cref{fig:lossyBinning} shows the relative loss of equi-width and equi-height quantization. The x-axis varies $\dv$ and the y-axis shows the mean absolute error: $ \textsc{MAE}(\mat{X}, \hat{\mat{X}}) = \sum_{1}^{|\mat{X}|}|x_i - \hat{x_i}|/ |\mat{X}|$. The upper and lower bounds of the blue and orange colored areas are the min/max absolute errors. Quantization and incurred errors show a linear relationship (log-scale plots) of roughly $\textsc{MAE}(\mat{X}, \Qv_\dv(\mat{X})) \approx 2\cdot\textsc{MAE}(\mat{X},\Qv_{2\dv}(\mat{X}))$, meaning if $\dv$ doubles, the MAE error is halved. Learned schemes can improve the MAE using higher $\dv$ or optimize for other goals such as model accuracy~\cite{ZhangYYH18}, compression size, or pareto-optimal combinations.

\subsection{Non-numerical Data}

Categorical values are commonly encoded with one-hot encoding, whereas text is often represented via word embeddings. Feature transformations producing numerical representations through binning, feature hashing, recoding, and one-hot encoding have the potential to compress encoded values. Some are lossy categorical transformations that reduce $d$. Feature hashing maps values to $\dv$ buckets, and for Natural Language Processing (NLP), one can limit the number of unique words or tokens ($d$) for encoding via lemmatization \cite{Dawson74} and stemming.
\begin{wrapfigure}{r}{8.5cm}
	\centering
	\includegraphics[scale=0.99]{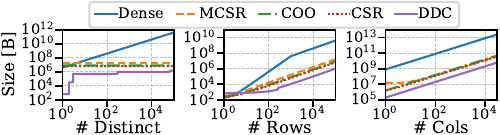}
	\caption{\label{fig:OneHotPotential}Output Memory Sizes of One-Hot/Dummy Coding an Input.}
	\Description{...}
\end{wrapfigure}
\cref{fig:OneHotPotential} shows the potential of compressing one-hot-encoded columns using dictionary compression compared to different sparse representations. The three sub-plots systematically vary the number of distinct values, rows, and columns of inputs (with base parameters \numprint{1000} distinct values, \numprint{100}k rows and 5 columns). The output shape is [\#rows, \#columns $\cdot$ $d$], and sparsity is $1/d$. The y-axis shows the in-memory size in bytes of the encoded output matrix. The dense allocations show worse performance in all cases beyond very few distinct values or rows. Sparse layouts such as Compressed Sparse Rows (CSR)~\cite{Bassiouni85}, COOrdinate matrix (COO)~\cite{Bassiouni85}, and Modified CSR (MCSR)~\cite{BoehmDEEMPRRSST16} yield good compression in all cases. Sparsity exploitation performs exceptionally well when scaling $d$. However, compared to a DDC compression~\cite{ElgoharyBHRR16,BaunsgaardB23, MullerCF14, WestmannKHM2000, AntoshenkovLM96}, all the other solutions allocate more memory. The dampened size increase for Dense in the middle plot is because the $d$ increases until 1000 and not above.

\subsection{Correlation}

\begin{wrapfigure}{r}{7.cm}
	\centering
	\includegraphics{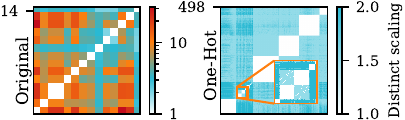}
	\caption{Relative $d$ Increase when Co-coding Features in Adult:
		Original and One-Hot Encoded Features.
		\label{fig:OneHotFeatureCorrelation}}
	\Description{...}
\end{wrapfigure}

Another property that impacts compressibility is the correlation between columns. \cref{fig:OneHotFeatureCorrelation} shows the relative increase in the number of distinct tuples when combining different columns in the Adult dataset (we removed one column with $d > 20k$). The left sub-figure shows the original features, while the right shows the one-hot encoded categorical features. Let $d_i$ be 
the number of distinct values in column $i$ and $d_{ij}$ the number of distinct tuples of co-coded columns $i$ and $j$. Then, each cell $c_{ij}$ in the figure shows $c_{ij} = 2d_{ij} / (d_i + d_j)$ to demonstrate the relative increase of distinct tuples by combining the columns. White (with $c_{ij} = 1$) indicates columns with perfect correlations, which, for instance, is the case for all pairs of one-hot-encoded columns originating from the same column. A greedy co-coding algorithm requires $\mathcal{O}(m^2)$ time (where $m$ is the number of columns) to discover these correlated columns. Ideally, co-coding would first group one-hot encoded features with perfect correlation and subsequently other correlated features. Rediscovering the correlation between columns is non-trivial and potentially very expensive since each combination of columns has to be analyzed (on a sample). The rediscovery is further complicated by ultra-sparse matrices and the existence of sparsity-exploiting compression schemes, where the full co-coding potential is often not analyzed in favor of fast compression. Interestingly, \cref{fig:OneHotFeatureCorrelation} shows a perfect correlation between the original features 3 and 4, while the one-hot encoded version does not perfectly co-code on all sub-combinations of those columns (see zoomed-in area). Instead, this perfect correlation is only detectable via evaluations of larger sub-groups. Therefore, pushing compression through feature transformations has the potential for both runtime reduction and improved compression.

\subsection{Pre-processing Time}

\begin{figure}[b]
	\centering
	\includegraphics[scale=.99]{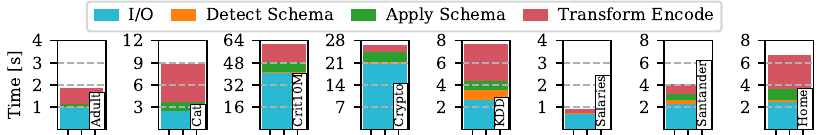}
	\vspace{-0.2cm}
	\caption{Breakdown of Reading, Detecting and Applying Schemas, and Losslessly Encoding Different Datasets. \label{fig:motivationTimeBreakdown}}
	\Description{...}
\end{figure}

\cref{fig:motivationTimeBreakdown} shows the execution time of pre-processing the different datasets. We read datasets in CSV format from disk, marked as I/O. In case of unknown data types, schema detection and application aim to specialize generic strings into integer and floating point data types where possible. We detect data types on a sample and apply them during data conversion. As a final step, the heterogeneous frame is transformed into matrices through \emph{transform encode}. There is potential to improve all these stages via compression. Reading compressed representations from disk reduces the number of read bytes. Frames saved with a schema can skip schema operations, but feature transformations are still required. AWARE~\cite{BaunsgaardB23} similarly observed---but did not exploit---that a combined pre-processing and compression analysis could significantly reduce the end-to-end execution time depending on the algorithm(s) and data used. We further extend this train of thought with new ideas on feature transformation techniques that can process compressed inputs directly without decompression to reduce execution time and memory consumption.

\section{Compressed Data Preparation}
\label{sec:frameCompression}

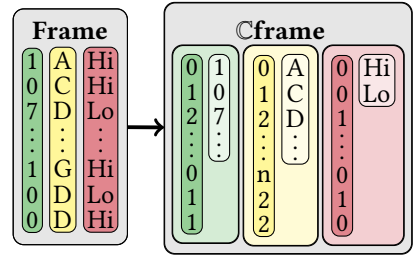
\begin{wrapfigure}{r}{5.48cm}
  \vspace{-0.45cm}
  \centering
  \begin{tikzpicture}[ampersand replacement=\&]
    \begin{scope}
      \matrix[frame, anchor=north](IN){
          1      \&[0.1cm] A         \&[0.1cm] Hi    \\
          0      \&        C         \&        Hi    \\
          7      \&        D         \&        Lo    \\[-0.2cm]
          \vdots \&        \vdots    \&        \vdots  \\
          1      \&        G         \&        Hi    \\
          0      \&        D         \&        Lo    \\
          0      \&        D         \&        Hi    \\
      };
      \node[mlabel](INL) at (IN.north){\textbf{Frame}};
      
    \end{scope}
    \node[bbCus, inner sep=0.mm, fit=(INL) (IN), minimum height = 3.1cm](INBB){};
    
    \begin{scope}[yshift=-0.1cm, xshift = 0.3cm]
       \begin{scope}[xshift=1.5cm]
          \matrix[frame, anchor=north](A){
              0      \&[0.1cm] 1         \\
              1      \&        0         \\
              2      \&        7         \\[-0.2cm]
              \vdots \&        \vdots    \\
              0      \&                  \\
              1      \&                  \\
              1      \&                  \\
          };
          
        \end{scope}
        \begin{scope}[xshift=2.5cm]
          \matrix[frame, anchor=north](B){
              0      \&[0.1cm] A         \\
              1      \&        C         \\
              2      \&        D         \\[-0.2cm]
              \vdots \&        \vdots    \\
              n      \&                  \\
              2      \&                  \\
              2      \&                  \\
          };
          
        \end{scope}
        \begin{scope}[xshift=3.6cm]
          \matrix[frame, anchor=north](C){
              0      \&[0.1cm] Hi        \\
              0      \&        Lo        \\
              1      \&                  \\[-0.2cm]
              \vdots \&                  \\
              0      \&                  \\
              1      \&                  \\
              0      \&                  \\
          };
          
        \end{scope}
        
        \node[bbCus, inner sep=0.mm, fit=(A), minimum height = 1.95cm](ABB){};
        \node[bbCus, inner sep=0.mm, fit=(B), minimum height = 1.95cm](BBB){};
        \node[bbCus, inner sep=0.mm, fit=(C), minimum height = 1.95cm](CBB){};
        
        \node[mlabel](CL) at ([yshift=0.1cm]BBB.north){$\mathbb{C}$\textbf{frame}};
         
    \end{scope}
       
    \node[bbCus, inner sep=1.mm, fit=(ABB) (BBB) (CBB) (CL)](COBB){};
     \begin{scope}[on background layer]
       \node[bbCus, inner sep=0.mm, fit=(INL) (IN), fill=black!10, minimum height = 3.1cm]{};
       \node[bbCus, thin, fit = (IN-1-1) (IN-7-1), inner sep = -0.15, rounded corners=1mm, fill=green!50]{};
       \node[bbCus, thin, fit = (IN-1-2) (IN-7-2), inner sep = -0.15, rounded corners=1mm, fill=yellow!50]{};
       \node[bbCus, thin, fit = (IN-1-3) (IN-7-3), inner sep = -0.15, rounded corners=1mm, fill=red!50]{};
       \node[bbCus, inner sep=0.mm, fit=(COBB), fill=black!10]{};
       \node[bbCus, inner sep=0.mm, fit=(ABB), fill=green!20]{};
       \node[bbCus, inner sep=0.mm, fit=(A-1-1) (A-7-1),rounded corners=1mm, fill=green!50](AMAPBB){};
       \node[bbCus, inner sep=0.mm, fit=(A-1-2) (A-4-2),rounded corners=1mm, fill=green!10](ADIBB){};
       \node[bbCus, inner sep=0.mm, fit=(BBB), fill=yellow!20]{};
       \node[bbCus, inner sep=0.mm, fit=(B-1-1) (B-7-1),rounded corners=1mm, fill=yellow!50](BMAPBB){};
       \node[bbCus, inner sep=0.mm, fit=(B-1-2) (B-4-2),rounded corners=1mm, fill=yellow!10](BDIBB){};
       \node[bbCus, inner sep=0.mm, fit=(CBB), fill=red!20]{};
       \node[bbCus, inner sep=0.mm, fit=(C-1-1) (C-7-1),rounded corners=1mm, fill=red!50](CMAPBB){};
       \node[bbCus, inner sep=0.mm, fit=(C-1-2) (C-2-2),rounded corners=1mm, fill=red!10](CDIBB){};
     \end{scope}
      \draw[->, line width = 0.5mm] (INBB) to (COBB);
  \end{tikzpicture}

  \vspace{-0.3cm}
  \caption{\label{fig:compressed_frame}The Compressed Frame Format.}
  \Description{...}
  \vspace{-0.9cm}
\end{wrapfigure}
This section describes BWARE's frame compression, compressed feature transformations to matrices, and compressed feature engineering.

\subsection{Compressed Frame Design}
\label{sec:CompressedFrameDesign}

Uncompressed frames are tables stored in columnar arrays. Each column can contain different value types. \cref{fig:compressed_frame} shows our $\mathbb{C}$frames using a dense dictionary compression (DDC) scheme per column. 
Each DDC column consists of a mapping array, $length = \#rows$, on the left, and a dictionary array, $length = d_i$ on the right. $d_i$ is $d$ in column $i$. The map contains value positions in the dictionary. 

\textbf{Compressed Size:} The compressed size depends on the number of distinct values, $d$, value type, and number of rows. The mapping supports 0 or 1 Bit and 1-, 2-, 3-, or 4 Byte encodings per value, $\#B$, (supporting up to 1, 2, 256, 64K, 16M, and 2G distinct values).
If $d_i$ is large, the dictionary allocation becomes too costly compared to the original value types, and we fall back to uncompressed arrays from the input frames columns.
We compress boolean columns, which increases the size but can be leveraged in BWARE's feature transformations and engineering.

\textbf{Type Conversion:} Type conversion improves allocations using specialized value types. We detect the value type on a sample of the data and fuse conversion and column compression. In case of casting errors, we re-detect a guaranteed correct value type and convert the column to the newly detected type. We support string, int, character, boolean, hexcode, and float types of different precision. The schema detection and application are critical because our system defaults to reading frames as strings unless a schema is provided on the initial read. For example, a hash encoded as a hex \texttt{"bcdef123"} but allocated as a string can be very costly.

\textbf{Simple Compression:} We do not co-code columns because many feature transformations use unique dictionaries for individual columns, and different columns can contain different value types. Subsequent workload-aware morphing (\cref{sec:morphing}) of the compressed format anyway tunes the final matrix compression with full support of a wide variety of different encoding schemes. The proposed transformation techniques would also work with other dictionary-based compression techniques (e.g., RLE~\cite{AbadiMF06}, SDC~\cite{BaunsgaardB23}, and OLE~\cite{ElgoharyBHRR16}), which we leave for future work.

\textbf{Compression:} We fuse type detection, type conversion, and DDC compression. For each column:
(1) We detect value types on a sample and try to apply the conversion while compressing. If the detected value type is invalid for the entire column, we revert to a full pass of the column to guarantee type detection.
(2) To compress, we allocate a \emph{mapping} that can encode as many unique IDs as rows.
(3) Then, we iterate through all column rows and build a hashmap to encode unique column values with contiguously increasing IDs. All values IDs are stored in the \emph{mapping}.
(4) We pack the \emph{mapping} into an improved format according to the $d_i$ elements encountered to improve $\#B$.
(5) We allocate a dictionary array, $D$,  and fill it by looping through the hashmap's key-value pairs $<k_i,v_i>$ and assign $D[v_i] = k_i$.
We do not compress the column, if the hashmap grows too large compared to the \#rows and detected value type. Even without DDC, type detection and conversion can still improve memory usage.

\textbf{Parallelization:} We na\"{\i}vely parallelize over all input columns because the compressed frame format independently compresses columns. However, some datasets contain few columns and many rows, and parallelizing only over columns does not fully utilize the available degree of parallelism. Therefore, each column thread further parallelizes the parsing of value types from strings---which can be costly (e.g., String to double~\cite{Lemire2021})---over row segments.

\setlength{\columnsep}{0.4cm}%

\subsection{Compressed Feature Transformations}
\label{sec:transformEncode}
\label{sec:CompressedFeatureTransformations}

\begin{wraptable}{r}{4.3cm}
   \vspace{-1.0cm}
   \centering
   \setlength\tabcolsep{2.5pt}
   \caption{Transform Encode Types.\label{tab:transformEncodeSupport}}
   \vspace{-0.4cm}
   \begin{tabular}{c|ccc}
      \toprule
      \textbf{Name}  & \textbf{OH} & $\mathbb{C}$\textbf{-In\&Out}  \\
      \midrule  
      Bin            & \checkmark     & \checkmark    \\
      Hash           & \checkmark     & \checkmark    \\
      Pass           & \checkmark     & \checkmark    \\
      Recode         & \checkmark     & \checkmark    \\
      Word Emb       &                & \checkmark    \\
      \bottomrule
   \end{tabular}
   \vspace{-0.4cm}
\end{wraptable}

Transform-encode encodes a heterogeneous frame into a homogeneous matrix by applying dedicated feature transformations (built-in function \texttt{transformencode()}). The operation produces two outputs: A matrix and a metadata frame to apply the same transformations to other frames. We support the transformations shown in \cref{tab:transformEncodeSupport}. Other numeric transformations can be subsequently performed in linear algebra (e.g., normalization). 

\textbf{Lossless:} We support two lossless transformations: \emph{Pass} returns the same values as the input cast to double. \emph{Pass} requires numeric inputs. \emph{Recode} encodes input values into contiguous integers for each unique value encountered (similar to DDC encoding, just throwing away the dictionary).

\textbf{Lossy:} Similarly, there are two lossy transformations: \emph{Bin} short for Binning, constructs $n$ buckets to encode the values into. The bins use equi-height or equi-width quantization. Equi-height constructs buckets with similar frequency of data points by calculating quantile boundaries. Equi-width extracts the minimum and maximum value and constructs buckets of equal ranges. The values returned from binning are bin IDs. Another technique, \emph{hash}, hashes each value and returns the hashed value modulo the maximum number of buckets to yield a bin ID. 

\textbf{One-hot Encoding:} One-hot encoding (\emph{OH})---which is also called \emph{dummy coding} (with subtle differences)---encodes contiguous integer values $i$ into one-hot sparse vector representations with a one set in cell $i$. This operation can be used on top of other transformations (which return integers) or on its own and thus, is supported in most cases, see \cref{tab:transformEncodeSupport}.

\textbf{Word Embeddings:} Encoding words into semantic-preserving numeric vectors is done via \emph{word embeddings}, which is a sequence of \emph{recoding}, \emph{one-hot encoding}, and matrix multiplication with an embedding matrix. $v$ denotes the size of each embedding vector. The matrix multiplication use a selection matrix (details in \cref{sec:selectionMatrixMultiply}), constructed via a contingency table on the recoded output.

\begin{wrapfigure}{r}{5.1cm}
	\vspace{-0.3cm}
	\centering
	\begin{tikzpicture}[ampersand replacement=\&, node distance= 1.25cm]
		\node[bbt1             ](F){\begin{tikzpicture}[ampersand replacement=\&]
    \node{
        \resizebox{1cm}{1cm}
        {
        \begin{tikzpicture}[ampersand replacement=\&]
            \matrix[Mtiles,
                column  1/.style={every node/.style={fulln, minimum width=1em, minimum height=4em, fill=green!50}},
                column  2/.style={every node/.style={fulln, minimum width=1em, minimum height=4em, fill=yellow!50}},
                column  3/.style={every node/.style={fulln, minimum width=1em, minimum height=4em, fill=red!50}},
             ](TILED){
                \  \& \  \& \ \\
            };
        \end{tikzpicture}
        }
    };
    \node at(0,0.7){Frame};
\end{tikzpicture}};
		\node[bbt1, right of=F ](CF){\begin{tikzpicture}[ampersand replacement=\&]
    \node{
        \resizebox{1cm}{1cm}
        {
        \begin{tikzpicture}[ampersand replacement=\&]
            \matrix[Mtiles,
                column  1/.style={every node/.style={fulln, minimum width=1em, minimum height=1em, fill=green!50}},
                column  2/.style={every node/.style={fulln, minimum width=1em, minimum height=4em, fill=yellow!50}},
                column  3/.style={every node/.style={fulln, minimum width=1em, minimum height=2em, fill=red!50}},
             ](TILED){
                \  \& \  \& \ \\
            };
        \end{tikzpicture}
        }
    };
    \node at(0,0.7){$\mathbb{C}$Frame};
\end{tikzpicture}};
		\node[bbt1, right of=CF](M){\input{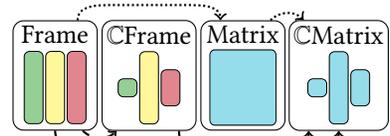}};
		\node[bbt1, right of=M , xshift = 0.025cm](CM){
\begin{tikzpicture}[ampersand replacement=\&]
    \node{
        \resizebox{1cm}{1cm}
        {
            \begin{tikzpicture}[ampersand replacement=\&]
                \matrix[Mtiles,
                column  1/.style={every node/.style={fulln, minimum width=1em, minimum height=1em, fill=blue!50}},
                column  2/.style={every node/.style={fulln, minimum width=1em, minimum height=4em, fill=blue!50}},
                column  3/.style={every node/.style={fulln, minimum width=1em, minimum height=2em, fill=blue!50}},
             ](TILED){
                \  \& \  \& \ \\
            };
            \end{tikzpicture}
        }
    };
    \node at(0,0.7){$\mathbb{C}$Matrix};
\end{tikzpicture}
};


		\draw[->, thick, densely dashed] ([xshift = -2mm]F.south east) 
		to[out = -45, in = -90-45] ([xshift = 2mm]CF.south west);

		\draw[->,thick, densely dashed] ([xshift=0.4cm]CF.south)
			to[out= -90, in= 180] ([xshift= 0.5cm, yshift=-0.2cm] CF.south)
			to[out= 0, in= 180]   ([xshift=-0.5cm, yshift=-0.2cm] CM.south)
			to[out= 0, in= - 90]  ([xshift=-0.4cm]CM.south);

		\draw[->,thick, densely dotted] ([xshift=0.3cm] F.north) 
			to[out= 90, in= 180]  ([xshift= 0.8cm, yshift=0.2cm] F.north)
			to[out= 0, in= 180]   ([xshift=-0.8cm, yshift=0.2cm] M.north)
			to[out= 0, in= 90]    ([xshift=-0.3cm]M.north);

		\draw[->, thick, densely dotted] ([xshift = -2mm]M.north east) to[out = 45, in = 90+45] ([xshift = 2mm]CM.north west);

		\draw[->,thick] ( F.south) 
			to[out= -90, in= 180] ([xshift= 0.5cm, yshift=-0.3cm] F.south)
			to[out= 0, in= 180]   ([xshift=-0.5cm, yshift=-0.3cm] CM.south)
			to[out= 0, in= -90]   (CM.south);

	\end{tikzpicture}
	\vspace{-0.45cm}
	\caption{Transform Encode Sequences.
	\label{fig:TransformCompressed}}
	\Description{...}
	\vspace{-0.3cm}
\end{wrapfigure}

\textbf{Compression Sequences:} \cref{fig:TransformCompressed} shows different transformation sequences. The abbreviations F-$\mathbb{C}$F is frame compression and M-$\mathbb{C}$M stands for matrix compression.

\textbf{Frame to Matrix (F-M-$\mathbb{C}$M):} The already existing baseline approach is to first transform-encode an uncompressed frame to an uncompressed matrix (F-M). Subsequently, the matrix is compressed (M-$\mathbb{C}$M) with existing lossless matrix compression techniques~\cite{BaunsgaardB23, FenganLYANPM19, ElgoharyBHRR16, FerraginaMGTKNST22}. However, the separate matrix compression has to extract statistics from the intermediate matrix again, many similar to the F-M transformation's statistics.

\textbf{Frame to Compressed Matrix (F-$\mathbb{C}$M):} The compression feature transformations are:
\begin{itemize}
   \item \textbf{Recoding:} Uses two passes: (1) construct a hashmap of unique values to continuous IDs, and (2) applying the assigned IDs. Finally, allocate a dictionary using the HashMap keys as values and values as offsets. \textbf{+Dummy:} use an identity matrix as a dictionary.
   \item \textbf{Pass-Through:} Takes a sample if uncompressed and verifies compressibility. If the column is incompressible, return an uncompressed column group. Otherwise, proceed as recode, but use the hashmap keys for the dictionary values.
   \item \textbf{Hashing:} Hashing does not need a hashmap. Instead, we directly allocate a dictionary similar to recode of k values and hash each tuple directly into the mapping. The hashing method may not use all buckets, potentially creating unnecessary entries in the compressed dictionary.
   \textbf{+Dummy:} Use an identity matrix of k rows and columns.
   \item \textbf{Bin:} Calculate the bin of each value and put it into the mapping. The dictionary is incrementing integers until $\dv$.
   \textbf{+Dummy:} Use an identity matrix of $\dv$ rows and columns.
\end{itemize}

\setlength{\columnsep}{0.6cm}%

\textbf{Compressed Frame to Compressed Matrix ($\mathbb{C}$F-$\mathbb{C}$M):} Compressed frame inputs offer multiple optimization opportunities. First, we skip constructing hashmaps by directly utilizing the dictionaries of a compressed frame. Second, we reuse the index structures (the map of DDC, but other structures for other compressions) allocated from the $\mathbb{C}$frame for the $\mathbb{C}$matrix. Compression is usually dominated by creating index structures because the ratio of distinct values is commonly small. Reusing the index structures makes the transformation scale in the number of distinct values rather than the number of rows. This approach is, however, only applicable if we use lossless transformations because lossy transformations have to reallocate or re-map their index structures.

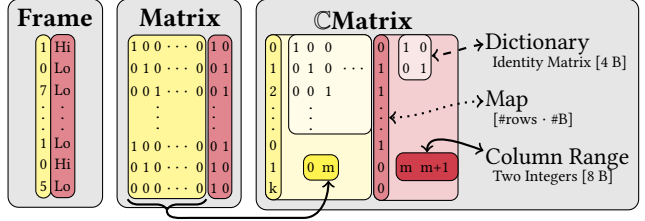
\begin{wrapfigure}{r}{8.2cm}
    \centering
    \vspace{-0.5cm}
    \begin{tikzpicture}
        \begin{scope}
            \matrix[frame, anchor = north, font = \tiny,
                minimum height=0.8em,](IN){
                1      &[0.00cm] Hi   \\[-0.039cm]
                0      & Lo  \\[-0.039cm]
                7      & Lo  \\[-0.2  cm]
                \vdots & \vdots \\[-0.039cm]
                1      & Lo \\[-0.039cm]
                0      & Hi   \\[-0.039cm]
                5      & Lo  \\[-0.039cm]
            };
            \node[](INL) at (0, 0.1cm){\textbf{Frame}};
            \begin{scope}[on background layer]
                \node[bbCus, inner sep=0.mm, fit=(INL) (IN), fill=black!10](INBB){};
                \node[bbCus, thin, fit = (IN-1-1) (IN-7-1), inner sep = -0.15, rounded corners=1mm, fill=yellow!50]{};
                \node[bbCus, thin, fit = (IN-1-2) (IN-7-2), inner sep = -0.15, rounded corners=1mm, fill=red!50]{};
            \end{scope}

        \end{scope}

        \begin{scope}[xshift=1.66cm]

            \matrix[frame,anchor = north,
            every node/.style={inner sep=0.3mm, minimum height=0.8em,
            font = \tiny},
            yshift =-0.005cm
            ](OUT){
                1 & 0 & 0      & $\cdots$ & 0 &[0.03cm] 1 & 0 \\[-0.015cm]
                0 & 1 & 0      & $\cdots$ & 0 & 0      & 1 \\[-0.013cm]
                0 & 0 & 1      & $\cdots$ & 0 & 0      & 1 \\[-0.217cm]
                  &   & \vdots &          &   & \vdots &   \\[0.045cm]
                1 & 0 & 0      & $\cdots$ & 0 & 0      & 1 \\[-0.034cm]
                0 & 1 & 0      & $\cdots$ & 0 & 1      & 0 \\[-0.034cm]
                0 & 0 & 0      & $\cdots$ & 0 & 1      & 0 \\
            };
            \node[](OUTL)at (0, 0.1cm){\textbf {Matrix}};

            \draw [thick,decorate,decoration={brace,amplitude=4pt,mirror,raise=0ex}](OUT-7-1.south west) -- (OUT-7-5.south east) ;

            \begin{scope}[on background layer]
                \node[xshift=0.1cm, yshift=0.16cm] at (OUT.south west)(TMP){};
                \node[bbCus, inner sep=0.mm, fit=(OUT.north east)  (TMP)  (OUTL), fill=black!10](OUTBB){};
                \node[bbCus, thin, fit = (OUT-1-1) (OUT-7-5), inner sep = -0.2, rounded corners=1mm, fill=yellow!50]{};
                \node[bbCus, thin, fit = (OUT-1-6) (OUT-7-7), inner sep = -0.2, rounded corners=1mm, fill=red!50]{};
            \end{scope}
        \end{scope}

        \begin{scope}[xshift=3.9cm]

            \matrix[frame,font = \tiny](COUT){
                0      & [0.1cm] 1 & 0     & 0  & & [0.03cm] & 0      & [0.1cm] 1  & 0 \\[-0.04cm]
                1      & 0   & 1      & 0  & $\cdots$   &  & 1      & 0            & 1 \\[-0.035cm]
                2      & 0   & 0      & 1  &            &  & 1      &              &   \\[-0.218cm]
                \vdots &     & \vdots &    &            &  & \vdots &              &   \\[-0.010cm]
                0      &     &        &    &            &  & 1      &              &   \\[-0.035cm]
                1      &     & 0      & m  &            &  & 0      & m            & \ \\[-0.105cm]
                k      &     &        &    &            &  & 0      &              &   \\[-0.04cm]
            };
            \node[font = \tiny](C2RE) at([xshift = 0.15cm, yshift = 0.065cm]COUT-6-9){m+1}; 

            \node[](COUTL)at (0.2cm, 0.07cm){$\mathbb{C}$\textbf{Matrix}};
            \begin{scope}[on background layer]
                \node[bbCus, thin, fit = (COUT-1-1) (COUT-7-1) (COUT-1-4), inner sep = -0.2, rounded corners=1mm, fill=yellow!50]{};
                \node[bbCus, thin, fit = (COUT-1-1) (COUT-7-1) , inner sep = -0.2, rounded corners=1mm, fill=yellow!50](C1M){};
                \node[bbCus, thin, fit = (COUT-1-2) (COUT-3-4) , inner sep = -0.2, rounded corners=1mm, fill=yellow!10](C1M){};
                \node[bbCus, thin, fit = (COUT-6-3) (COUT-6-4) , inner sep = -0.2, rounded corners=1mm, fill=yellow!80](C1M){};
                \node[bbCus, thin, fit = (COUT-1-7) (COUT-6-7) (COUT-1-9) (C2RE), inner sep = -0.2, rounded corners=1mm, fill=red!50]{};
                \node[bbCus, thin, fit = (COUT-1-7) (COUT-7-7), inner sep = -0.2, rounded corners=1mm, fill=red!50](C2M){};
                \node[bbCus, thin, fit = (COUT-1-8) (COUT-2-9), inner sep = -0.2, rounded corners=1mm, fill=red!10](C2D){};
                \node[bbCus, thin, fit = (COUT-6-8) (C2RE), inner sep = -0.2, rounded corners=1mm, fill=red!80](C2R){};

            \end{scope}

            \node[xshift=1.7cm, yshift=-0.25cm, anchor = west, align = left](D){\small Dictionary};
            \node[xshift=1.8cm, yshift=-0.5cm, anchor = west, align = left]{\tiny Identity Matrix [4 B]};
            \node[xshift=1.7cm, yshift=-1.0cm, anchor = west, align = left](M){\small Map};
            \node[xshift=1.8cm, yshift=-1.25cm, anchor = west, align = left]{\tiny [\#rows $\cdot$ \#B]};
            \node[xshift=1.7cm, yshift=-1.75cm, anchor = west, align = left](R){\small Column Range};
            \node[xshift=1.8cm, yshift=-2.00cm, anchor = west, align = left]{\tiny Two Integers [8 B]};

            \draw[<->, thick, dashed] ([xshift = 0.1cm, yshift = 0.0cm]D.west) to[out=188, in=5] (C2D);
            \draw[<->, thick, dotted] ([xshift = 0.1cm, yshift = 0.0cm]M.west) to[out=180, in=0] (C2M);
            \draw[<->, thick] ([xshift = 0.1cm, yshift = 0.1cm]R.west) to[out=160, in=90] (C2R.north);
            

            \draw[->, thick]([xshift=-0.175cm]OUT.south) to[out= -90, in= -180] 
                ([xshift=0.12cm, yshift=-0.1cm]OUT.south) -- 
                ([xshift=1.645cm, yshift=-0.1cm]OUT.south) 
                to[out =0, in=-90] (COUT-6-4.south);

            \begin{scope}[on background layer]
                \node[color= black!10](TMP2) at (0, 0.1cm){Frame}; 
                \node[bbCus, inner sep=0.mm, fit=(COUT) (TMP2)(D)(M)(R), fill=black!10](COUTBB){};
                \node[bbCus, thin, fit = (COUT-1-1) (COUT-7-1) (COUT-1-4) (COUT-2-5), inner sep = -0.2, rounded corners=1mm, fill=yellow!20]{};
                \node[bbCus, thin, fit = (COUT-1-1) (COUT-7-1) , inner sep = -0.2, rounded corners=1mm, fill=yellow!50](C1M){};
                \node[bbCus, thin, fit = (COUT-1-2) (COUT-3-4) (COUT-2-5) (COUT-4-3) , inner sep = -0.2, rounded corners=1mm, fill=yellow!10](C1D){};
                \node[bbCus, thin, fit = (COUT-6-3) (COUT-6-4) , inner sep = -0.2, rounded corners=1mm, fill=yellow!80](C1R){};
                \node[bbCus, thin, fit = (COUT-1-7) (COUT-7-7) (COUT-1-9) (C2RE) , inner sep = -0.2, rounded corners=1mm, fill=red!20]{};
                \node[bbCus, thin, fit = (COUT-1-7) (COUT-7-7), inner sep = -0.2, rounded corners=1mm, fill=red!50](C2M){};
                \node[bbCus, thin, fit = (COUT-1-8) (COUT-2-9), inner sep = -0.2, rounded corners=1mm, fill=red!10](C2D){};
                \node[bbCus, thin, fit = (COUT-6-8) (C2RE), inner sep = -0.2, rounded corners=1mm, fill=red!80](C2R){};
            \end{scope}
        \end{scope}
    \end{tikzpicture}
    \vspace{-0.8cm}
    \caption{\label{fig:transformEncodeEx} Recode and Dummy-code Two Columns.}
    \Description{...}

\end{wrapfigure}

\textbf{Example:}{\textit{ \cref{fig:transformEncodeEx} shows recoding and one-hot-encoding a frame of two columns. The uncompressed and compressed matrix results are shown in the middle and on the right. The transformation returns $d_i$ matrix columns for each input column. The unique values are incrementally mapped to encoded values. The mapping from row indexes to dictionary entries is the value's recoded IDs. The dictionary is a virtual identity matrix (stored in a single integer). Each column group contains a column range with a start and end index. The mapping size depends on the number of rows and $d_i$. Assuming $|\mat{X}| = $\numprint{1000} and $d_1 = 200$, the left mapping uses 1 B/row and the right column uses 1 b/row. The $\mathbb{C}$matrix then requires $1032 + 176 =1208$ B plus object/pointer overheads. If the input frame is compressed, like in \cref{fig:compressed_frame}, the output can point to the input mapping of the compressed frame column. In such cases, the time complexity is $\mathcal{O}(1)$ instead of $\mathcal{O}(n)$. Furthermore, the allocation becomes constant with $20$ B of pointers.}}

\begin{wrapfigure}{r}{8.2cm}
    \vspace{-0.2cm}
    \centering
    \begin{tikzpicture}
        \begin{scope}[xshift=-1cm]
            \matrix[frame, anchor = north](IN){
                1      &[0.00cm] \\
                0      &  \\
                7      &  \\[-0.2cm]
                \vdots &  \\
                5      &  \\
            };
            \node[](INL) at (0, 0.07cm){\textbf{Frame}};
            \node(Size) at (0.5,-0.2) {\fontsize{4.9pt}{4.9pt}\selectfont $ \in \mathbb{N}^{|\mat{X}|}$};

            \node[rotate=90](INNAME) at ([xshift=-0.2cm]IN-3-1.west){\fontsize{5pt}{5pt}\selectfont Tokenized Vector};
            \begin{scope}[on background layer]
                \node[bbCus, inner sep=0.mm, fit=(INL) (IN) (Size), fill=black!10](INBB){};
                \node[bbCus, thin, fit = (IN-1-1) (IN-5-1), inner sep = -0.15, rounded corners=1mm, fill=red!50](INMAPBB){};
            \end{scope}

        \end{scope}

        \begin{scope}[xshift=2.cm]

            \matrix[frame](COUT){
                1      & [0.1cm] 1  & 0 & 0 &                \\[0.02cm]
                0      & 0   & 1      & 0  & $\cdots$        \\[-0.015cm]
                7      & 0   & 0      & 1  &                 \\[-0.198cm]
                \vdots &     & \vdots &    &                 \\[0.015cm]
                5      &   0 &  m     &    &                 \\
            };

            \node[](COUTL)at (0.0cm, 0.07cm){\textbf {$\mathbb{C}$Matrix}};
            \node at (0.45,-1.75) {\fontsize{4.9pt}{4.9pt}\selectfont $ \in \mathbb{N}^{|\mat{X}|\texttt{x}m}$};

            \begin{scope}[on background layer]
                    \node[bbCus, inner sep=0.mm, 
                    fit=(COUTL) (COUT), fill=black!10](COUTBB){};
                \node[bbCus, thin, fit = (COUT-1-1) (COUT-5-1) (COUT-1-4) (COUT-2-5) (COUT-5-3), inner sep = 1, rounded corners=1mm, fill=yellow!30](CMAPBB){};
                \node[bbCus, thin, fit = (COUT-1-1) (COUT-5-1) , inner sep = -0.2, rounded corners=1mm, fill=red!50](C1M){};
                \node[bbCus, thin, fit = (COUT-1-2) (COUT-3-4) (COUT-4-3) (COUT-2-5) , inner sep = -0.2, rounded corners=1mm, fill=yellow!10](C1D){};
                \node[bbCus, thin, fit = (COUT-5-2) (COUT-5-3) , inner sep = -0.2, rounded corners=1mm, fill=yellow!80](C1C){};

            \end{scope}

        \end{scope}

        \draw[->, thick](INBB.east) -- ([ yshift=0.04cm]COUTBB.west);
        \node[] at ([xshift=0.55cm, yshift=0.2cm]INBB.east){\fontsize{7pt}{7pt}\selectfont Encode};
        \node[] at ([xshift=0.55cm, yshift=0.4cm]INBB.east){\fontsize{7pt}{7pt}\selectfont One-Hot};

        \draw[<->,  dashed]
               ([yshift= -0.45cm]C1M.west) 
            -- ([yshift= -0.5cm]INMAPBB.east);
        \node[] at ([xshift=0.55cm, yshift=-0.54cm]INBB.east){\fontsize{5pt}{5pt}\selectfont Pointer or};
        \node[] at ([xshift=0.55cm, yshift=-0.74cm]INBB.east){\fontsize{5pt}{5pt}\selectfont Reallocation};
        \node[] at ([xshift=0.55cm, yshift=-0.93cm]INBB.east){\fontsize{5pt}{5pt}\selectfont of Mapping};

        \begin{scope}[xshift=3.9cm, yshift= -0.4cm]
    
			\node[fulln,
                    fill=orange!50,
					minimum height = 1cm,
					minimum width = 0.8cm](W){\ };
		
            \node at (0, 0.2) {$\mat{W}$};
            \node at (0,-0.2) {\fontsize{5pt}{5pt}\selectfont $ \in \mathbb{R}^{m\texttt{x}v}$};
		    \node at(0,0.6){\fontsize{5pt}{5pt}\selectfont Embedding};

            \node[fulln,
                fill=purple!30,
                minimum height = 0.6cm,
                minimum width = 0.9cm](MM) at (0, -1){\ };

            \node[yshift=0.1cm] at(MM){\fontsize{6pt}{6pt}\selectfont Matrix};
            \node[yshift=-0.1cm] at(MM){\fontsize{6pt}{6pt}\selectfont Multiply};

            \draw[->, thick] ([yshift=-0.58cm]COUTBB.east) to[out=0, in =180] (MM);
            \draw[->, thick] (W.south) to[out=-90, in =90] (MM);

        \end{scope}

        \begin{scope}[xshift= 5.8cm, yshift = -0.24cm]

            \matrix[frame](EC){
                \node[pointer, fill=red!50,yshift=-0.1cm](EC-1-1){};   & \node[pointer, fill=orange!50](EC-1-2){};   \\[0.2cm]
                    0 &  v  \\
            };

            \node[](ECL)at (0.0cm, 0.08cm){\textbf {$\mathbb{C}$Emb}};
            \node(ESIZE) at (0.1,-1.2) {\fontsize{5pt}{5pt}\selectfont $ \in \mathbb{R}^{|\mat{X}|\texttt{x}v}$};

            \begin{scope}[on background layer]
                \node[bbCus, inner sep=0.mm, 
                    fit=(EC) (ECL) (ESIZE), fill=black!10](ECBB){};
                \node[bbCus, thin, fit = (EC-1-1) (EC-1-2) (EC-2-1) (EC-2-2) , inner sep = 0.7, rounded corners=1mm, fill=green!30](ECBBI){};
                \node[bbCus, thin, fit = (EC-2-1) (EC-2-2) , inner sep = -0.2, rounded corners=1mm, fill=green!80](ECD){};
            \end{scope}

            \draw [->, thick] (MM) to[out = 0, in = 180] (ECBB);

            \draw [<->, dotted] 
                (EC-1-2.center) to[in =0, out = 150] 
                ([xshift=-0.4cm, yshift = -0.05cm]EC-1-2.north west) to[in = 0, out = 180] 
                ([]W.east);

            \node[]at ([xshift=-1.1cm, yshift = 0.20cm]EC-1-2.north west){\fontsize{5pt}{5pt}
                \selectfont Dictionary};
            \node[]at ([xshift=-1.1cm, yshift = 0.05cm]EC-1-2.north west){\fontsize{5pt}{5pt}
                \selectfont Pointer};

            \draw [<->, dashed] 
                (EC-1-1.center) to[in =90, out = 200] 
                ([xshift=-0.2cm, yshift = -0.65cm]EC-1-1.north west) to[in = 90, out = -90] 
                ([xshift=-0.2cm, yshift = -1.05cm]EC-1-1.north west) to[in = 0, out = -90]
                ([xshift=-1.2cm, yshift = -1.55cm]EC-1-1.north west) to[in = 0, out =180] 
                ([xshift=-3.7cm, yshift = -1.55cm]EC-1-1.north west) to[in = -45, out =180]  
                ([xshift=-0.04cm, yshift=-0.8cm]C1M.east);

            \node[]at ([xshift=-0.8cm, yshift = -1.225cm]EC-1-1.north west){\fontsize{5pt}{5pt}      
                \selectfont Mapping};
            \node[]at ([xshift=-0.8cm, yshift = -1.375cm]EC-1-1.north west){\fontsize{5pt}{5pt}      
                \selectfont Pointer};
        \end{scope}

    \end{tikzpicture}
    \vspace{-0.3cm}
    \caption{\label{fig:transformWordEmb}Compressed Linear Algebra Word Embedding.}
    \Description{...}
    \vspace{-0.2cm}
\end{wrapfigure}
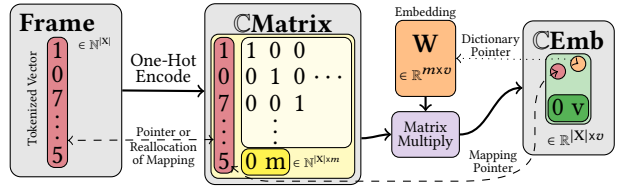

\textbf{Compressed Word Embedding:} \cref{fig:transformWordEmb} shows how we perform a compressed word embedding for a single column input in $\mathcal{O}(1)$, only requiring shallow copies of (i.e., pointers to) already allocated intermediates. Since the embedding is a right matrix multiply and the intermediate compressed matrix's dictionary is an identity matrix, the embedding multiplication constructs a new compressed result with a pointer to the full embedding matrix, reused as the dictionary of the dictionary encoding.

   
\begin{wraptable}{r}{8.3cm}
   \centering
   \setlength\tabcolsep{3pt}
   \vspace{-0.2cm}
   \caption{\label{tab:transformEncodeSize}Transform-Encode Column Asymptotic Size.}
   \vspace{-0.2cm}

   \begin{tabular}{c|c|c|c}
      \toprule
         & \textbf{F-M} & \textbf{F-$\mathbb{C}$M}  & \textbf{$\mathbb{C}$F-$\mathbb{C}$M} \\
      \midrule 
      Recode \& Pass       & $8|\mat{X}|$    & $\#B|\mat{X}|+8d_i$  & $constant$  \\
      Bin \& Hash    & $8|\mat{X}|$    & $\#B|\mat{X}|+8\dv$ & $\#B|\mat{X}|+8\dv$  \\
      \midrule
      \multicolumn{4}{c}{With One-Hot / Dummy-Coding} \\
      \midrule
      Recode \& Pass & $12|\mat{X}|$    & $\#B|\mat{X}|$  & $constant$  \\
      Bin \& Hash    & $12|\mat{X}|$    & $\#B|\mat{X}|$  & $\#B|\mat{X}|$  \\
      \midrule
      Word Embed & $(8v+12)|\mat{X}|$ &  $\#B|\mat{X}|$ & constant \\
      \bottomrule
   \end{tabular}
   \vspace{-0.2cm}
\end{wraptable}

\textbf{Intermediate Sizes:} \cref{tab:transformEncodeSize} shows the sizes formulas of outputs. \textbf{F-M} is the uncompressed standard transformation, while \textbf{F-$\mathbb{C}$M} and \textbf{$\mathbb{C}$F-$\mathbb{C}$M} produce compressed matrices. The first half of the table is without one-hot encoding. The one-hot \textbf{F-M} size assumes CSR output. Otherwise, dense representations require $8|X|d_i$. '$constant$' means the compressed input index structure is used directly in the output.

\subsection{Compressed Feature Engineering}
\label{sec:augmentation}
\label{sec:CompressedFeatureEngineering}

Feature engineering constructs or modifies features to improve the accuracy of ML models.

\textbf{Modified Features:} We define feature modifications as element-wise operations that change all instances of a feature equally. A typical operation is normalization. Normalization has multiple forms such as scale and shift or min-max scaling. These modifications are needed for well-behaved training of ML algorithms. Elementwise operations are supported in compressed space by prior work~\cite{ FerraginaMGTKNST22, FenganLYANPM19, BaunsgaardB23, ElgoharyBHRR16}. In essence, we similarly perform most of the modifications with time complexity in the number of distinct items, $\mathcal{O}(d_i)$ for each column group.

\textbf{Additional Features:} One can add features by joining on keys or construct new from existing features. An example is to expand $\mat{X}$ with $\mat{X}'= \text{cbind}(\mat{X}, \mat{X}^2)$, where we concatenate \mat{X} and its squared representation. Such features allow simple linear models to take non-linearities into account, known as the kernel trick \cite{Aizerman1964}. The append, however, requires allocating the full extended matrix.  In contrast, BWARE combines new column groups with minimal additional allocations.
\begin{wrapfigure}{r}{8.4cm}

   \vspace{-0.2cm}
   \centering
   \begin{tikzpicture}[node distance=0.35cm, ampersand replacement=\&]
      \begin{scope}
         \matrix[frame, anchor = north](A){
                0  \& [0.1cm] 7.0  \\
                1  \&         7.1  \\ 
                0  \&         9.3  \\ 
                2  \&              \\
                1  \&              \\
                2  \&         0    \\
            };

         \node[mlabel](AL) at (A.north){{\large \textbf{$\mat{X}$}}};
      \end{scope}
     
      \begin{scope}[xshift=2.5cm]
         \matrix[frame, anchor = north](B){
                  \& [-0.1cm] 49.00  \\
                  \&         50.41  \\ 
                  \&         86.49  \\ 
                \   \&                \\
                \   \&                \\
                  \&  [0.1cm]       0      \\
            };
            \node[bbCus, inner sep=1.5mm, outer sep=1mm,
               fill=blue!50, rectangle](BId) at
               ([yshift = -0.0cm, xshift= 0.05cm]B-5-1) {};
         \node[mlabel](BL) at (B.north){\large \textbf{$\mat{X}^2$}};
      \end{scope}

      \begin{scope}[xshift=6.5cm]
         \matrix[frame, anchor = north](N){
               7.0 \& 49.00 \\
                      7.1 \& 50.41 \\ 
                      9.3 \& 86.49 \\ 
            \                  \&  \\
            \ \                \&  \\
                        \& 0,1  \\
        };

        \node[bbCus, inner sep=1.5mm, outer sep=1mm,
         fill=blue!50, rectangle](NId) at
        ([yshift = -0.0cm, xshift= 0.05cm]N-5-1) {};
        \node[mlabel](NL) at (N.north){\large \textbf{$\mat{X}'$}};
       
      \end{scope}

      \node[bbCus, inner sep=0.mm, fit=(A)  (AL)](ABB){}; 
      \node[bbCus, inner sep=0.mm, fit=(B)  (BL)](BBB){}; 

      \begin{scope}[on background layer]
         \node[bbCus, inner sep=0.mm, fit=(A)  (AL), fill=black!10](ABB){};
         \node[bbCus, thin, fit = (A-1-1) (A-6-1),  
           inner sep = -0.2, rounded corners=1mm, fill=blue!50](AID){};
         \node[bbCus, thin, fit = (A-1-2) (A-3-2),  
           inner sep = -0.2, rounded corners=1mm, fill=blue!10](ADIC){};
         \node[bbCus, thin, fit = (A-6-2),  
           inner sep = -0.2, rounded corners=1mm, fill=blue!80](ACOL){};
     \end{scope}

      \begin{scope}[on background layer]
         \node[bbCus, inner sep=0.mm, fit=(B)  (BL), fill=black!10](BBB){};
         \node[bbCus, thin, fit = (B-1-2) (B-3-2),  
           inner sep = -0.2, rounded corners=1mm, fill=blue!10](BDIC){};
         \node[bbCus, thin, fit = (B-6-2),  
           inner sep = -0.2, rounded corners=1mm, fill=blue!80](BCOL){};
      \end{scope}

      \begin{scope}[on background layer]
         \node[bbCus, inner sep=0.mm, fit=(N)  (NL), fill=black!10](NBB){};
         \node[bbCus, thin, fit = (N-1-1) (N-3-2),  
            inner sep = -0.2, rounded corners=1mm, fill=green!10](NDIC){};
         \node[bbCus, thin, fit = (N-6-2),  
            inner sep = -0.2, rounded corners=1mm, fill=green!80](NCOL){};
      \end{scope}

      \draw[->, thick, dotted] ([xshift=0.145cm, yshift=-0.145cm]BId.north west) 
         to[out=135, in=0] ([xshift=2cm]A-4-1) -- (A-4-1);
      \draw[->, thick, dotted] ([xshift=0.145cm, yshift=-0.145cm]NId.north west) 
         to[out=135, in=0] ([xshift=5cm]A-4-1) -- (A-4-1);

      \draw[->, thick, dotted] (B-6-2.west) 
         to[out=180, in=0] (A-6-2);

      \node at([xshift=2.1cm, yshift=0.2cm]BId.north west){\tiny Index Pointer };
      \node at([xshift=-1.1cm, yshift=0.1cm]B-6-2.west){\tiny Columns };
      \node at([xshift=-1.1cm, yshift=-0.1cm]B-6-2.west){\tiny Pointer };
      
      \node at([xshift=0.7cm, yshift=0.4cm]ABB.east)(Op1){$\square^2$};
      \node at([xshift=1.3cm, yshift=0.4cm]BBB.east)(Op2){$cbind(\square,\square)$};
      
      \begin{scope}[on background layer]
         \node[bbCus, inner sep=0.mm, fit=(Op1), fill=red!20](Op1BB){};
         \node[bbCus, inner sep=0.mm, fit=(Op2), fill=red!20](Op2BB){};
      \end{scope}

      \draw[->,thick] (ADIC) to[out=0, in=180]  (Op1BB);
      \draw[->,thick] (Op1BB) to[out=0, in=180] (BDIC);

      \draw[->,thick] ([yshift=-0.3cm]ADIC.east) 
         to[out=0, in=180] ([yshift=-0.05cm, xshift=-0.8cm]BDIC.south)
         to[out=0, in=180] ([yshift=-0.05cm]BDIC.south)
         to[out=0, in=180] ([yshift=-0.05cm, xshift=0.4cm]BDIC.south)
         to[out=0, in=180] ([yshift=-0.1cm]Op2BB.west);
      \draw[->,thick] (BDIC) to[out=0, in=180] ([yshift=0.1cm]Op2BB.west);
      \draw[->,thick] (Op2BB) to[out=0, in=180] (NDIC);

      \end{tikzpicture}
   \vspace{-0.4cm}
   \caption{\label{fig:augment}Extending Features in a Co-coded Column Group.}
   \Description{...}

   \vspace{-0.2cm}
\end{wrapfigure}
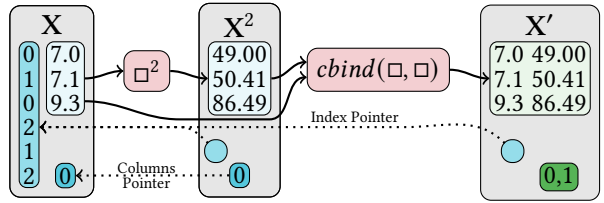
\cref{fig:augment} shows the modification on a DDC encoding (but we support this operation for multiple compressed encoding types). The first step performs the scalar power operation. Scalar power is a dictionary-only operation, reducing the allocation to a new dictionary and maintaining pointers back to the original input mapping. When performing the column bind (\emph{cbind}) operation, we detect that both indexes point to the same mapping, which indicates perfect correlation, and thus allows the direct combination into a co-coded column group. Similar exploitation strategies exist when subsets of columns are modified and appended. Furthermore, the compressed append can add multiple non-linearities at the same time $\mat{X}'' = \text{cbind}(\mat{X}, \mat{X}^2, \text{log}(\mat{X}), \sqrt{\mat{X}})$.

\textbf{Performance:} $|\mat{X}| / d_i$ defines the potential speedup of compressed feature engineering of individual columns because the new features have perfect correlation with the original features and can share index structures. Exploiting the co-coding removes redundant compression analysis of the augmented matrices. Many compressed operations benefit from extensive co-coding. For example, left matrix multiplication (LMM), with compressed right and uncompressed left inputs, benefits because pre-aggregation is independent of the number of co-coded columns.

\section{Morphing}
\label{sec:morphing}

\def\bottom#1#2{\hbox{\vbox to #1{\vfill\hbox{#2}}}}
\tikzset{samplerow/.style={bbCus, minimum height = 0.1cm, minimum width=1.7cm, rounded corners=0mm, fill = black}}
\tikzset{samplestat/.style={bbCus, thin, minimum height = 0.5cm, minimum width=0.3cm}}
\tikzset{samplestatn/.style={ inner sep=-0.5mm,thin, minimum height = 0.5cm, minimum width=0.3cm}}

\begin{wrapfigure}{r}{8.45cm}
    \vspace{-0.75cm}
    \centering
        \begin{tikzpicture}[node distance=0.35cm, ampersand replacement=\&]

            \begin{scope}[yshift=0.2cm, xshift = -0.2cm]
                \node[bbt1](IM){\input{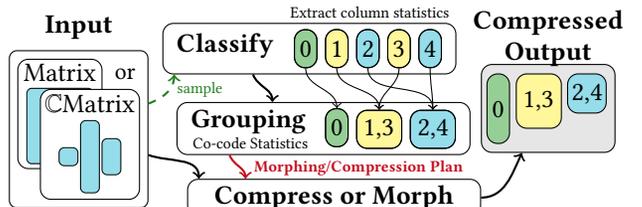}};
                \node[bbt1, xshift=0.4cm, yshift=-0.4cm](CIM){
\begin{tikzpicture}[ampersand replacement=\&]
    \node{
        \resizebox{1cm}{1cm}
        {
            \begin{tikzpicture}[ampersand replacement=\&]
                \matrix[Mtiles,
                column  1/.style={every node/.style={fulln, minimum width=1em, minimum height=1em, fill=blue!50}},
                column  2/.style={every node/.style={fulln, minimum width=1em, minimum height=4em, fill=blue!50}},
                column  3/.style={every node/.style={fulln, minimum width=1em, minimum height=2em, fill=blue!50}},
             ](TILED){
                \  \& \  \& \ \\
            };
            \end{tikzpicture}
        }
    };
    \node at(0,0.7){$\mathbb{C}$Matrix};
\end{tikzpicture}
};

                \node[anchor = north, font = \small]at([xshift = 0.3cm]IM.north east){or};

                \begin{scope}[on background layer]
                    \node[bbCus, fit=(IM) (CIM), inner sep = 3, rounded corners=1mm](InBB){};
                \end{scope}
                \node[mlabel,text width= 2cm, text centered](IL) at ([yshift=0.2cm]InBB.north){\textbf{Input}};
            \end{scope}

            \begin{scope}[yshift=-0.2cm]
                \begin{scope}[xshift=2cm, yshift=1.2cm] 
                    \node[mlabel](SL) {\textbf{Classify}};
                    \matrix[cg, yshift=0.44cm,
                        every node/.style={samplestat}]
                    (ST) at (1.9cm, 0){
                        0 \& [0.1cm] 1 \& [0.1cm] 2 \& [0.1cm] 3 \& [0.1cm] 4  \\
                    };

                    \node[mlabel](CD)at (ST.north){\fontsize{6}{6}\selectfont{ Extract column statistics}};

                   

                    \begin{scope}[on background layer]

                        \node[samplestat, fit=(ST-1-1),fill = green!50  ]{};
                        \node[samplestat, fit=(ST-1-2),fill = yellow!50]{};
                        \node[samplestat, fit=(ST-1-3),fill = blue!50  ]{};
                        \node[samplestat, fit=(ST-1-4),fill = yellow!50]{};
                        \node[samplestat, fit=(ST-1-5),fill = blue!50 ]{};
                        \node[bbCus, fit=(ST) (SL), minimum width = 3.9cm](CLBB){};

                    \end{scope}

                \end{scope}

                \begin{scope}[xshift= 2.31cm, yshift = 0.2cm] 

                    \node[mlabel](GL){\textbf{Grouping}};
                    \matrix[cg, yshift=0.4cm,
                        every node/.style={samplestat},
                        column 2/.style={nodes={minimum width=0.6cm, text width = 0.6cm, align=center}},
                        column 3/.style={nodes={minimum width=0.6cm, text width = 0.6cm, align=center}}
                    ]
                    (GST) at (1.89cm, 0){
                        0 \& [0.1cm] 1,3 \& [0.1cm] 2,4 \\
                    };

                    \node[mlabel](GD)at ([yshift=-0.1cm]GL.south){\fontsize{6}{6}\selectfont{ Co-code Statistics}};

                    \draw[->](ST-1-1.south) to[out=-90, in= 90] (GST-1-1.north);
                    \draw[->](ST-1-2.south) to[out=-90, in= 90] (GST-1-2.north);
                    \draw[->](ST-1-3.south) to[out=-90, in= 90] (GST-1-3.north);
                    \draw[->](ST-1-4.south) to[out=-90, in= 90] (GST-1-2.north);
                    \draw[->](ST-1-5.south) to[out=-90, in= 90] (GST-1-3.north);

                    \begin{scope}[on background layer]

                        \node[samplestat, fit=(GST-1-1),fill = green!50  ]{};
                        \node[samplestat, fit=(GST-1-2),fill = yellow!50, minimum width=0.6cm]{};
                        \node[samplestat, fit=(GST-1-3),fill = blue!50  , minimum width=0.6cm ]{};
                        \node[bbCus, fit=(GST) (GL), minimum width = 3.9cm](GSBB){};

                    \end{scope}

                    \draw[->, thick]([xshift=-.75cm]CLBB.south) to[out = -70, in = 90]([xshift=-.68cm, ]GSBB.north);
                \end{scope}

                \begin{scope}[xshift=3.45cm, yshift=-.8cm] 
                    \node[mlabel](CO){\textbf{Compress or Morph}};

                    \begin{scope}[on background layer]

                        \node[bbCus, fit= (CO), minimum width = 3.9cm](COBB){};

                    \end{scope}
                    \draw[color = red,->, thick]([xshift=-1.25cm]GSBB.south) to[out = -70, in = 90]([xshift=-1.18cm, ]COBB.north);

                \end{scope}

            \end{scope}
           
            \draw[color = green!80!black,->, dashed, thick](InBB) to[out=20, in=-90]([xshift=-1.8cm]CLBB.south);
            \node[color = green!80!black,font = \tiny]at([yshift = -0.2cm, xshift = 0.5cm]CLBB.south west){sample};
            \draw[->, thick](InBB) to[out=-20, in=90]([xshift=-1.8cm]COBB.north);

            \node[]at([yshift=0.15cm, xshift=0.3cm]COBB.north){\fontsize{6}{6}\selectfont{ \color{BerlinRed} \textbf{Morphing/Compression Plan}}};

            \begin{scope}[xshift= 6.3cm, yshift=.5cm] 
                \matrix[cg, yshift=0.4cm,
                    every node/.style={samplestatn},
                    column 1/.style={nodes={anchor=north, minimum height = 1.cm, }},
                    column 2/.style={nodes={anchor=north, minimum height = .8cm, 
                        minimum width=0.6cm, text width = 0.6cm, align=center}},
                    column 3/.style={nodes={anchor=north, minimum height = .6cm, 
                        minimum width=0.6cm, text width = 0.6cm, align=center}},
                    ]
                    (C){
                        0 \& [0.09cm] 1,3 \& [0.04cm] 2,4  \\
                        };

                \node[mlabel,text width= 2cm, text centered](CL) at ([yshift=0.2cm]C.north){\textbf{Compressed Output}};

                \begin{scope}[on background layer]

                    \node[bbCus, fit=(C), fill = black!10](CBB){};
                    \node[samplestat, fit=(C-1-1),fill = green!50  ]{};
                    \node[samplestat, fit=(C-1-2),fill = yellow!50, minimum width=0.6cm]{};
                    \node[samplestat, fit=(C-1-3),fill = blue!50  ]{};

                \end{scope}

            \end{scope}

            \draw[->, thick](COBB.east) to[out=0, in=-115]([xshift=-0.35cm]CBB.south);

        \end{tikzpicture}
    
    \vspace{-0.45cm}
    \caption{\label{fig:CompressionSequence}Morphing or Compression Sequence.}
    \Description{...}
    \vspace{-0.1cm}
\end{wrapfigure}

Our novel morphing-based compression transforms uncompressed dense or sparse, as well as compressed matrices into tuned compressed matrix representations. The morphing sequence is shown in \cref{fig:CompressionSequence}. First, for uncompressed inputs, we extract column statistics from the input and group columns according to these statistics and workload. Second, in case of a compressed matrix input, we directly reuse the statistics of pre-existing co-coded columns and skip unnecessary exploration. The result is a morphing/compression plan that contains a recipe for which columns to merge and what type of encoding to use.

\begin{figure}[t]
    \centering
    \input{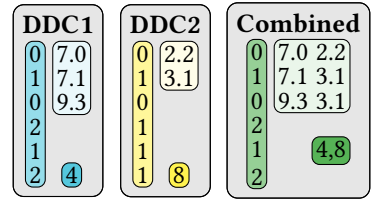}
    \begin{minipage}{5.2cm}
\begin{figure}[H]
   \begin{tikzpicture}[node distance=0.35cm, ampersand replacement=\&]
      \begin{scope}[yshift = -0.25mm]
         \matrix[frame, anchor = north, minimum height = 1.85cm](A){
                0  \& [0.1cm] 7.0  \\
                1  \&         7.1  \\ 
                0  \&         9.3  \\ 
                2  \&              \\
                1  \&              \\
                2  \&         4    \\
            };

         \node[mlabel, minimum height=0.5cm](AL) at (A.north){\textbf{DDC1}};
      \end{scope}

      \begin{scope}[xshift=1.43cm, yshift = -0.25mm]
         \matrix[frame, anchor = north, minimum height = 1.85cm](B){
                0  \& [0.1cm] 2.2  \\
                1  \&         3.1  \\ 
                0  \&              \\ 
                1  \&              \\
                1  \&              \\
                1  \&         8    \\
            };

         \node[mlabel, minimum height=0.5cm](BL) at (B.north){\textbf{DDC2}};

      \end{scope}


       

      \begin{scope}[xshift=3.2cm]
         \matrix[frame, anchor = north, minimum height = 2.05cm](O){
            0  \& [0.1cm] 7.0  \&  2.2 \\
            1  \&         7.1  \&  3.1 \\ 
            0  \&         9.3  \&  3.1 \\ 
            2  \&              \&      \\
            1  \&              \&  4,8 \\
            2  \&              \&   \\
        };

        \node[mlabel, minimum height=0.5cm](OL) at (O.north){\textbf{Combined}};
       
      \end{scope}

      \node[bbCus, inner sep=0.mm, fit=(A) (AL)](ABB){}; 
      \node[bbCus, inner sep=0.mm, fit=(B)  (BL)](BBB){}; 
      \node[bbCus, inner sep=0.mm, fit=(O) (OL)](OBB){}; 


      \begin{scope}[on background layer]
         \node[bbCus, inner sep=0.mm, fit=(A)  (AL), fill=black!10](ABB){};
         \node[bbCus, thin, fit = (A-1-1) (A-6-1),  
           inner sep = -0.2, rounded corners=1mm, fill=blue!50](AID){};
         \node[bbCus, thin, fit = (A-1-2) (A-3-2),  
           inner sep = -0.2, rounded corners=1mm, fill=blue!10](ADIC){};
         \node[bbCus, thin, fit = (A-6-2),  
           inner sep = -0.2, rounded corners=1mm, fill=blue!80](ACOL){};
     \end{scope}

      \begin{scope}[on background layer]
         \node[bbCus, inner sep=0.mm, fit=(B)  (BL), fill=black!10](BBB){};
         \node[bbCus, thin, fit = (B-1-1) (B-6-1),  
           inner sep = -0.2, rounded corners=1mm, fill=yellow!50](BID){};
         \node[bbCus, thin, fit = (B-1-2) (B-2-2),  
           inner sep = -0.2, rounded corners=1mm, fill=yellow!10](BDIC){};
         \node[bbCus, thin, fit = (B-6-2),  
           inner sep = -0.2, rounded corners=1mm, fill=yellow!80](BCOL){};
      \end{scope}

      \begin{scope}[on background layer]
         \node[bbCus, inner sep=0.mm, fit=(O)  (OL), fill=black!10](OBB){};
         \node[bbCus, thin, fit = (O-1-1) (O-6-1),  
           inner sep = -0.2, rounded corners=1mm, fill=green!50](OID){};
         \node[bbCus, thin, fit = (O-1-2) (O-3-3),  
           inner sep = -0.2, rounded corners=1mm, fill=green!10](ODIC){};
         \node[bbCus, thin, fit = (O-5-3),  
           inner sep = -0.2, rounded corners=1mm, fill=green!80](OCOL){};
       \end{scope}

   \end{tikzpicture}
   \vspace{-0.25cm}
   \caption{\label{fig:combine_morph}Combining Two DDC Groups}
   \Description{...}
\end{figure}

\end{minipage}
\end{figure}

\textbf{Morphed Combining of Compressed Columns:} To avoid decompression, we designed a co-coding algorithm that takes two encoded columns and returns a compressed co-coded column. We support combining various input column encodings. \cref{alg:morphingCombine} and \cref{fig:combine_morph} show the combination of DDC column groups. The two column-groups to combine are on the left and the combined output on the right. A na\"{\i}ve combination would produce the cartesian product of the dictionaries. Instead, our solution materialize only dictionary tuples that co-appear in the index structures. The cbind example from the previous section is a form of combining, where the dictionary's number of unique tuples does not grow. Each cell value in a na\"{\i}ve mapping can be calculated via $i_R = i_1 + i_2d_1$. Instead of populating the combined index with the na\"{\i}ve index values, we indirectly populate a hashmap to assign the combined index. Each combined dictionary tuple can then be looked up through the hashmap. The asymptotic runtime of creating such co-coded column groups is $\mathcal{O}(|\mat{X}|+d_R)$, where typically $d_R \ll |\mat{X}|$.

\textbf{Morphing a Column Encoding:} Combined column groups might not be in the correct encoding per the morphing plan. Therefore, the final step is to morph individual groups into other encoding types. Since most of our encodings are variations of dictionary encoding, the conversion is simple. In general, we try to change encodings while reusing intermediates as much as possible. In practice, changing encodings typically only change the index structure while keeping dictionaries.

\textbf{Fallback Morphing Execution:} Sometimes, the set of column groups selected for co-coding use heterogeneous encoding schemes, making it hard to have specialized combining algorithms for all. The fallback solution---in case specialized kernels are non-existent for combinations of encodings---is to decompress the selected morphing columns into a temporary matrix followed by a standard compression from scratch. The fallback case is often avoided because the \texttt{transformencode} currently only uses DDC. This fallback allocates a potentially expensive uncompressed matrix of size \#row and \#columns to combine. We have specialized methods for most permutations of SDC, DDC, CONST, EMPTY and Uncompressed column groups, therefore, avoiding the fallback.

\section{Compressed I/O}
\label{sec:io}

Prior work on workload-aware lossless matrix compression used online compression after local or distributed reads, where---at least for local compression---the entire uncompressed input matrix needed to fit in memory. This approach restricts the size of compressible matrices and redundantly compresses the input matrix for every program execution. To address these limitations, we extended our BWARE compression framework to read and write compressed blocks and support continuous compression of streams (collections) of blocks.

\subsection{Compressed Data Layout}

\begin{figure}[t]
    \begin{minipage}{0.49\linewidth}
        \begin{figure}[H]

		\centering
		\begin{tikzpicture}
				\begin{scope}[yshift = -0.cm]
						\matrix[tiles,
								row 1 column 1/.style={every node/.style={fulln, minimum width=3em,minimum height=4em}},
						](IN){
								M \\
						};
						\draw [thick,decorate,decoration={brace,amplitude=4pt,mirror,raise=0ex}]
						([xshift=-0.03cm]IN-1-1.north east) -- ([xshift=0.03cm]IN-1-1.north west) ;
						\draw [thick,decorate,decoration={brace,amplitude=4pt,mirror,raise=0ex}]
						([yshift=-0.03cm]IN-1-1.north west) -- ([yshift=0.03cm]IN-1-1.south west) ;
						\node at ([yshift=0.3cm]IN-1-1.north){$k$};
						\node at ([xshift=-0.3cm]IN-1-1.west){$i$};

						\node at ([xshift=0.62cm, yshift=0.18cm]IN-1-1.east){Tile};

						\node at ([yshift=-0.24cm]IN.south){\small In-Memory};
				\end{scope}

				\begin{scope}[xshift = 2.1cm, yshift = -0.15em]
						\matrix[tiles,
								column  2/.style={every node/.style={fulln, minimum width=0.6em, fill=green!50}},
								row  3/.style={every node/.style={fulln, minimum height=0.6em,fill=red!50}},
								row 3 column 2/.style={every node/.style={fulln, minimum width=0.6em,minimum height=0.6em, fill=black!50}},
						](TILED){
								\  & \   \\
								\  & \   \\
								\  & \   \\
						};

						\draw [thick,decorate,decoration={brace,amplitude=4pt,mirror,raise=0ex}]
						([xshift=-0.03cm]TILED-1-1.north east) -- ([xshift=0.03cm]TILED-1-1.north west) ;
						\draw [thick,decorate,decoration={brace,amplitude=4pt,mirror,raise=0ex}]
						([yshift=-0.03cm]TILED-1-1.north west) -- ([yshift=0.03cm]TILED-1-1.south west) ;

						\node at ([yshift=0.3cm, xshift=-0.3cm]TILED-1-1.north west)(BS){$B$};
						\draw[->, thick]([xshift=-4pt]TILED-1-1.west) to[out=180, in=-90] (BS);
						\draw[->, thick]([yshift=4pt]TILED-1-1.north) to[out=90, in=0] (BS);

						\draw [thick,decorate,decoration={brace,amplitude=4pt,mirror,raise=0ex}]
						([xshift=-0.03cm]TILED-1-2.north east) -- ([xshift=0.03cm]TILED-1-2.north west) ;
						\node at ([yshift=0.3cm]TILED-1-2.north)(BSe){\tiny$k$ \% $B$};

						\draw [thick,decorate,decoration={brace,amplitude=4pt,mirror,raise=0ex}]
						([yshift=-0.03cm]TILED-3-1.north west) -- ([yshift=0.03cm]TILED-3-1.south west) ;
						\node at ([xshift=-0.5cm]TILED-3-1.west)(BSe){\tiny$i$ \% $B$};

						\node at ([yshift=-0.25cm]TILED.south){\small Tiled Blocks};

						\node at ([xshift=0.5cm, yshift=0.18cm]TILED.east){I/O};
						\node[fit=(TILED)](TILEDBB){};
				\end{scope}

				\begin{scope}[xshift = 4.3cm, yshift = 0.35cm]
						\matrix[cluster](DIS){
								\ \\
								\ \\
								\ \\
						};
						\node at ([yshift=-0.15cm]DIS.south){\small Disk Partitions};


						\matrix[smalltiles,
								column  1/.style={every node/.style={smalln, minimum width=0.6em, fill=green!50}},
								column  2/.style={every node/.style={smalln, minimum height=0.6em,fill=red!50, yshift=-0.13cm}},
								yshift=0.4cm] at ([yshift=-0.025cm]DIS-1-1)(N1){
								\  & \   \\
						};

						\matrix[smalltiles,
								column  2/.style={every node/.style={smalln, minimum width=0.6em, fill=green!50}},
								yshift=0.4cm] at ([yshift=-0.025cm]DIS-2-1)(N2){
									\  & \  \\
									};

						\matrix[smalltiles,
									column 2/.style={every node/.style={smalln, minimum width=0.6em,minimum height=0.6em, fill=black!50, yshift=-0.13cm}},
								yshift=0.4cm] at ([yshift=-0.025cm]DIS-3-1)(N3){
								\  & \  \\
						};

						\node[fit=(N1)(N2)(N3)(DIS)](SPARKBB){};

				\end{scope}

				\draw[<->, thick]([xshift=-0.1cm]TILEDBB.east) -- ([xshift= 0.1cm, yshift=-0.14cm]SPARKBB.west);

				\draw[<->, thick](IN) to[out=0, in=180] (TILED);

		\end{tikzpicture}
		\vspace{-0.4cm}
		\caption{\label{fig:tiledFormat} Uncompressed Tiled Format.}
		\Description{NONE}
\end{figure}
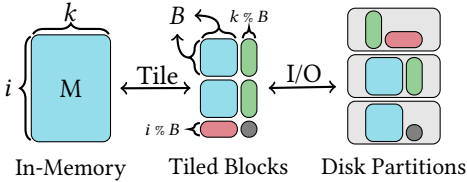
    \end{minipage}
    \begin{minipage}{0.49\linewidth}
        \tikzset{indexS/.style ={
	inner sep=0.1mm,
	minimum height=2.3em,
	minimum width=.1em,
	fill=black!90, draw,
	outer sep=0.1mm,
	rounded corners=0.2mm
}}

\tikzset{indexSDisk/.style ={
	inner sep=0.1mm,
	minimum height=1.3em,
	minimum width=.1em,
	fill=black!90, draw,
	outer sep=0.1mm,
	rounded corners=0.1mm
}}

\tikzset{indexEs/.style ={
	matrix of nodes, anchor=north,
	every node/.style={indexS},
	row sep=.7em,
	column sep=0.2em
}}

\tikzset{indexEsDisk/.style ={
	matrix of nodes, anchor=north,
	every node/.style={indexSDisk},
	row sep=.7em,
	column sep=0.2em
}}

\tikzset{smalln/.style={ fulln,
minimum height=1.8em,
minimum width=1.8em}}

\tikzset{smalltiles/.style={tiles,
every node/.style={smalln},
			}}

\begin{figure}[H]
\centering
\vspace{0.06cm}

\begin{tikzpicture}
\begin{scope}
	\matrix[indexEs,
		yshift=-0.5cm, 
	](IDX_IN){
			\ & \ & \  \\
	};

	\matrix[tiles,
			column  1/.style={every node/.style={fulln, minimum height=0.7em, fill=yellow!50}},
			column 2/.style={every node/.style={fulln,  minimum width=0.6em, minimum height=1.em, fill=orange!50}},
			column 3/.style={every node/.style={fulln,  minimum width=0.6em, minimum height = 0.55em, fill=purple!50}},
		yshift=0.2cm
	](DICT_IN){
			\ & \ & \ \\
	};

	\begin{scope}[on background layer]
		\node[bbCus, inner sep=0.mm, fit=(IDX_IN) (DICT_IN), minimum height = 1.95cm, fill= black!10](CBB){};
	\end{scope}

	\node[anchor= east](INDEXESLABEL) at ([xshift=1.3cm, yshift=0.3cm]IDX_IN.east){\tiny Indexes};
	\node[anchor= east](DICTSLABEL)        at ([xshift=1.3cm, yshift=0.0cm]DICT_IN.east){\tiny Dictionaries};

	\draw[->, dashed]([yshift=-0.1cm]INDEXESLABEL.north)
		to[out = 130, in = 0] ([yshift=0.4cm, xshift=-0.05cm]IDX_IN.east);

	\draw[->, dashed]([yshift=0.1cm]DICTSLABEL.south)
		to[out = -130, in = 0] ([yshift=-0.1cm, xshift=-0.05cm]DICT_IN.east);

	\node at ([xshift=0.4cm, yshift=-0.38cm]CBB.east){Tile};

	\node at ([yshift=-0.2cm]CBB.south){\small  $\mathbb{C}$ In-Memory};
\end{scope}

\begin{scope}[xshift= 2.3cm, yshift = -0.2cm]
	\matrix[indexEs,
		row 1/.style={every node/.style={indexS,minimum height=0.8em}},
		row 2/.style={every node/.style={indexS,minimum height=0.8em}},
		row 3/.style={every node/.style={indexS,minimum height=0.4em}},
		column 2/.style={column sep=.7em}
	](IDX_SEP){
			\ & \ & \  \\
			\ & \ & \  \\
			\ & \ & \  \\
	};

	\draw [thick,decorate,decoration={brace,amplitude=4pt,mirror,raise=0ex}]
	([xshift=-0.1cm,yshift= 0.03cm]IDX_SEP-1-1.north west) --
	([xshift=-0.1cm,yshift=-0.03cm]IDX_SEP-1-1.south west) ;

	\node at ([xshift=-0.45cm]IDX_SEP-1-1.west)(BS){$B$};

	\begin{scope}[on background layer]
		
		\node[bbCus, inner sep=0.8mm, rounded corners=1.mm, fit=(IDX_SEP-1-1) (IDX_SEP-1-2), fill= blue!50](){};
		\node[bbCus, inner sep=0.8mm, rounded corners=1.mm, fit=(IDX_SEP-2-1) (IDX_SEP-2-2), fill= blue!50](){};
		\node[bbCus, inner sep=0.8mm, rounded corners=1.mm, fit=(IDX_SEP-3-1) (IDX_SEP-3-2), fill= blue!50](){};
		\node[bbCus, inner sep=0.8mm, rounded corners=1.mm, fit=(IDX_SEP-1-1) (IDX_SEP-1-2), fill= blue!50](){};
		\node[bbCus, inner sep=0.8mm, rounded corners=1.mm, fit=(IDX_SEP-2-1) (IDX_SEP-2-2), fill= blue!50](){};
		\node[bbCus, inner sep=0.8mm, rounded corners=1.mm, fit=(IDX_SEP-3-1) (IDX_SEP-3-2), fill= blue!50](){};
		\node[bbCus, inner sep=0.8mm, rounded corners=1.mm, fit=(IDX_SEP-1-3), fill= blue!50](){};
		\node[bbCus, inner sep=0.8mm, rounded corners=1.mm, fit=(IDX_SEP-2-3), fill= blue!50](){};
		\node[bbCus, inner sep=0.8mm, rounded corners=1.mm, fit=(IDX_SEP-3-3), fill= blue!50](){};
		\node[bbCus, inner sep=0.8mm, rounded corners=1.mm, fit=(IDX_SEP-1-3), fill= blue!50](){};
		\node[bbCus, inner sep=0.8mm, rounded corners=1.mm, fit=(IDX_SEP-2-3), fill= blue!50](){};
		\node[bbCus, inner sep=0.8mm, rounded corners=1.mm, fit=(IDX_SEP-3-3), fill= blue!50](){};
		\node[bbCus, inner sep=0.0mm, fit=(IDX_SEP), minimum height = 1.95cm,draw=none](IndexBB){};
	\end{scope}

	\node at ([yshift=0.1cm]IndexBB.south){\small Tiled Indexes};

	\draw[<->, thick]([yshift=0.3cm]IDX_IN.south east)
		--([xshift = 1cm,yshift=0.3cm]IDX_IN.south east)
		to[out = 0, in = 180] (IndexBB.west);

\end{scope}

\begin{scope}[xshift = 4.5cm, yshift = 0.38cm]

\begin{scope}[on background layer]
	\matrix[cluster](DIS){
			\ \\
			\ \\
			\ \\
	};
\end{scope}

	\matrix[smalltiles, minimum width = 7em,
		column  1/.style={every node/.style={smalln, minimum height=0.7em, fill=yellow!50}},
		column  2/.style={every node/.style={smalln, minimum width=0.6em, minimum height=1.em, fill=orange!50}},
		column  3/.style={every node/.style={smalln, minimum width=0.6em, minimum height = 0.55em,fill=purple!50}},
		yshift=0.3cm] at ([xshift= -0.0cm]DIS-1-1)(N1){
		\  & \  & \ \\
	};

	\matrix[indexEsDisk, minimum width = 7em,
		row 1/.style={every node/.style={indexS, minimum height=0.8em}},
		row 1 column 2/.style={column sep=.7em},
		row 1 column 4/.style={column sep=.7em},
		row 1 column 5/.style={column sep=.7em, every node/.style={indexS, minimum height=0.4em}},
		yshift=0.27cm] at ([xshift= -0.0cm]DIS-2-1)(N2){
		\  & \  & \  & \ & \ \\
	};

	\matrix[indexEsDisk, minimum width = 7em,
		row 1/.style={every node/.style={indexS, minimum height=0.8em}},
		row 1 column 1/.style={column sep=.7em},
		row 1 column 2/.style={column sep=.7em},
		row 1 column 3/.style={every node/.style={indexS, minimum height=.4em}},
		row 1 column 4/.style={every node/.style={indexS, minimum height=.4em}},
		yshift=0.27cm] at ([xshift= -0.0cm]DIS-3-1)(N3){
			\  & \  & \ & \ \\
	};

\begin{scope}[on background layer]
	\node[bbCus, inner sep=0.8mm, rounded corners=1.mm, fit=(N3-1-1), fill= blue!50](){};
	\node[bbCus, inner sep=0.8mm, rounded corners=1.mm, fit=(N3-1-2), fill= blue!50](){};
	\node[bbCus, inner sep=0.8mm, rounded corners=1.mm, fit=(N3-1-3)(N3-1-4), fill= blue!50](){};

	\node[bbCus, inner sep=0.8mm, rounded corners=1.mm, fit=(N2-1-1)(N2-1-2), fill= blue!50](){};
	\node[bbCus, inner sep=0.8mm, rounded corners=1.mm, fit=(N2-1-3)(N2-1-4), fill= blue!50](){};
	\node[bbCus, inner sep=0.8mm, rounded corners=1.mm, fit=(N2-1-5), fill= blue!50](){};
	\node[bbCus, fit= (DICT_IN), fill= blue!10](){};
	\node[bbCus, fit= (IDX_IN), fill= blue!50](){};
\end{scope}

\node at ([xshift=0.55cm, yshift=0.18cm]IndexBB.east){I/O};
\draw[<->, thick]([yshift=-0.2cm]IndexBB.east) to[out = 0, in = 180] ([yshift=-0.3cm]DIS.west);
\end{scope}

\node at ([yshift=-0.15cm]DIS.south){\small Disk Partitions};
\draw[<->, thick]([yshift=0.16cm, xshift=-0.05cm]DICT_IN.east) 
	to[out = 0, in= 180](DIS-1-1.west);

\end{tikzpicture}
\vspace{-0.4cm}
\caption{\label{fig:CompressedTiledFormat} Compressed Tiled Format.}
\Description{NONE}
\end{figure}
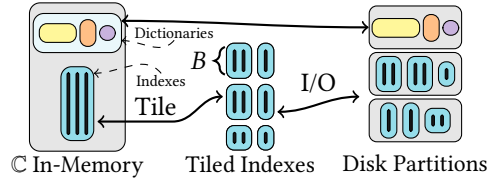
    \end{minipage}
\end{figure}

The on-disk format is a tiled format allowing distributed reads of collections of pairs of index and matrix blocks. \cref{fig:tiledFormat} shows the structure. The format excels in reading from local and distributed (HDFS) storage since multiple partitions can be read in parallel~\cite{DeadG08_MapReduce}. 
Reading and writing the matrix/frame formats have unique challenges to avoid decompression. When reading, we combine multiple, potentially differently compressed, blocks. When writing, we have to tile our index structures. For both read and write, we support local and distributed operations.

\textbf{Partitions \& Tiles:} Partitions hold multiple tiles and are written to individual files. The minimum number of tiles in each partition is determined by the partition sizes. We use minimum partition sizes of 128MB in HDFS (default block size) and 16KB in local (largest common disk block size). Partitions are allowed to grow larger than the minimum size.

\textbf{Compressed Disk Format:} The compressed format on disk can be seen in \cref{fig:CompressedTiledFormat}. We split the compressed index structures and the dictionaries into separate partition files. For local writes, we write the dictionaries only once (into an optional dictionary file) and rely on the read logic to combine the indexes and dictionaries again (e.g., via Spark \texttt{broadcast} for distributed reads). However, blocks compressed independently in distributed operations save index structure and dictionary together similarly to the uncompressed formats. The difference in behavior avoids collecting the distributed compressed representations to deduplicate the dictionaries but reduces the compression ratio because of duplicate dictionaries.

\subsection{Reading and Writing Compressed Data}

\textbf{Local Reading:} When reading a compressed matrix to local memory, we combine all the blocks into one consolidated columnar compression scheme. Such a read happens when reading a compressed input from disk, but is also applicable when collecting outputs of Spark operations. The simplest case of combining is if all blocks in a column use the same compression scheme. In this case, only one of the dictionaries is collected, and each index structure can be combined directly. In other cases, it is more complicated and might lead to decompression or morphing of individual sub-blocks and re-compression. Morphing enables changing the compression scheme of sub-blocks into the consolidated scheme without decompression. However, selected group encodings have methods to combine directly with other types such as \texttt{CONST}, \texttt{EMPTY} and \texttt{DDC}.

\textbf{Distributed Reading:} To read a compressed matrix in Spark, we construct a sequence of RDD (resilient distributed dataset) operations that lazily evaluate and materialize the compressed sub-blocks. If no separate dictionary file exists, all tiles should be self-contained (with index and dictionaries). In that case, we return a \emph{PairRDD} of indexes and blocks. If there is a dictionary file, we first read the index structures into \emph{PairRDDs} of matrix indexes and compressed index blocks and another of column index and dictionary blocks. We then join the dictionaries with the index structures to construct the combined compressed blocks. The result is a collection of fully self-contained compressed blocks. If the dictionaries grow above the tile size, then the compression makes each distributed block larger than uncompressed blocks. This fallback does not happen in practice because before writing any compressed block (or sub-block), we check if an uncompressed dense or sparse version is smaller and use the smallest.

\begin{algorithm}[!t]
    \caption{Serialized DDC Fused Update And Encode.}\label{alg:SerializedDDCCompression}
    \small
    \begin{algorithmic}
        \Require{$Matrix: \mat{X}$, Scheme: $S$;  $C \gets S_{Cols}$: , $M \gets S_{Map}$, $D \gets S_{Dict}$}

        \State $m_s \gets M.\textsc{size}()$ \Comment{{\color{commentgreen}\textbf{HashMap size before updates}}}
        \State $I_R \gets  \textsc{constructIndex}(\textsc{rows}(\mat{X}))$ \Comment{{\color{commentgreen}\textbf{Allocate output index}}}
        \For{ $r$ in $\mat{X}.\textsc{rows}()$} 
            \State $t \gets \textsc{extractTuple}(r, \mat{X}, C)$ \Comment{{\color{commentgreen}\textbf{Extract tuple from row}}}
            \State $i \gets \mat{M}\textsc{.putIfAbsent}(t, M.\textsc{size}())$ \Comment{{\color{commentgreen}\textbf{Increment size on new}}}
            \If{$I_R.\textsc{notValid}(i)$} $\textsc{Fail}$ \Comment{{\color{commentgreen}\textbf{Check support of value}}}
            \Else{ $I_R[r] \gets i$}
            \EndIf
        \EndFor
        \If{$m_s < M.\textsc{size}()$} $D \gets \textsc{updateDict}(M, D)$ 
        \EndIf

        \noindent
        \Return $\textsc{DDCColGroup}(I_R, D, C)$
    \end{algorithmic}
\end{algorithm}

\textbf{Update \& Encode:} To support large matrices or streaming use cases (e.g., continuous data collection), we can apply a compression plan on a stream of continuously arriving matrix blocks. The technique allows dynamic updates to a scheme to compress matrices without analysis, and each scheme can be applied in parallel. \cref{alg:SerializedDDCCompression} shows the DDC encoding variant (we support seven encodings in total). Given a compression scheme (determined from a sample), we first try to use a fused compression kernel that performs one pass over the input block. We use a one-pass algorithm because we are usually memory-bandwidth-bound. We start by allocating the output index structure and remembering the starting number of distinct values. The loop extracts the value tuples from the matrix depending on what columns are encoded. Then, we probe the map, if the tuple is not contained, we assign a new incremented ID, otherwise use an existing ID. If we encounter many new distinct tuples, the index structure might be unable to encode them. In such cases, we abort and fall back to a two-pass algorithm that first updates and then encodes in two loops. If the map size is equal to the beginning, no new values are encountered and we reuse the previous materialized dictionary. Otherwise, we update the dictionary. A benefit of this scheme is that \emph{all} previously encoded blocks can use the latest dictionary for computations.

\subsection{New Compressed Operations}

Slicing continuous ranges of rows from compressed frames and matrices is important for mini-batch algorithms, pre-processing, and writing. Additionally, data-centric ML pipelines often sample or select rows from input datasets via so-called selection matrix multiplications. In order to support these operations, BWARE provides dedicated compressed operations.

\textbf{Slicing Ranges:} Extracting continuous row sections from the compressed data is cheap by slicing the index structures and keeping a pointer to the input's dictionaries. We use this approach to tile compressed matrices for I/O, broadcasting to Spark, and linear algebra range selections.

\label{sec:selectionMatrixMultiply}
\textbf{Selection Matrix Multiply:} Selecting a random sample from a matrix can be done via matrix multiplication. If we define a sparse matrix $\mat{S}$, with a single 1 in each row and then left matrix multiply it with a matrix $\mat{X}$, this operation selects rows from $\mat{X}$. A 1's position in $\mat{S}$ determines the source row by its column position and the target row by its row position. This multiplication is, for instance, used in the initialization of K-Means where $k$ random rows for each run in K-Means act as cluster starting points. For compressed selection multiplication, we use a left compressed matrix multiply that does not pre-aggregate the intermediate matrix, unlike GCM~\cite{FerraginaMGTKNST22}, TOC~\cite{FenganLYANPM19}, AWARE~\cite{BaunsgaardB23}, and CLA~\cite{ElgoharyBHRR16}. Instead, we use the non-zero values of the left side matrix (guaranteed to be all 1) to selectively extract compressed row tuples, and decompress them into the output matrix. This solution leverages the index structures of the compression schemes.

\newpage

\setlength{\columnsep}{0.6cm}%

\section{Compiler and Runtime Integration}
\label{sec:compiler}
\input{fig/systemBase.tex}

\noindent
Data-centric ML pipelines transform data through multiple stages from disk
over pre-processing and augmentation to ML algorithms. \cref{fig:MLPipeline} shows a pipeline containing nested loops for finding the optimal pre-processing primitives. The stages comprise reading \nodeLabel{1}, a loop for different feature transformations specs $t$ \nodeLabel{2}, a loop for augmentation strategies $a$ \nodeLabel{3}, and the training loop of an algorithm \nodeLabel{4}, exemplified with a conjugate-gradient linear regression. There is potential to exploit redundancy via the previously presented techniques. However, adding compression to the stages would require hand-tuning the individual compound techniques based on the transformations, augmentations, and algorithms used. Instead, we propose an optimizing compiler and runtime that dynamically introduces the compression primitives in given linear algebra programs.

\textbf{Compiler:} We decide, at compile time, where to inject compression and morphing instructions. User-defined linear algebra programs, such as \cref{fig:MLPipeline}, are first compiled into a hierarchy of statement blocks (for conditional control flow and function calls) containing directed acyclic graphs (DAG) of high-level operations (HOPs) per last-level statement block. Each HOP is compiled to one or more low-level operators (LOPs). We detect HOPs with morphing potential by considered operations such as \texttt{read} and \texttt{transformencode}, but also operations like rounding (e.g., \texttt{floor}) or comparisons (e.g., $<=$), which produce integer and boolean outputs. For each candidate HOPs, we then construct a workload vector of affected, data-dependent operations summarizing the workload costs of the respective intermediate. If the workload summary indicates potential for improvement, the HOP is marked, appending a morphing LOP to its compiled sequence of LOPs.

\textbf{Runtime:} The runtime morphing has access to the compile-time workload vectors, allowing us to adapt the matrix to the workload. Since morphing supports compressed and uncompressed inputs, we handle unforeseen circumstances (e.g., after conditional data modification), adjusting the compression while still adhering to subsequent workload and data characteristics.

\begin{figure}[H]
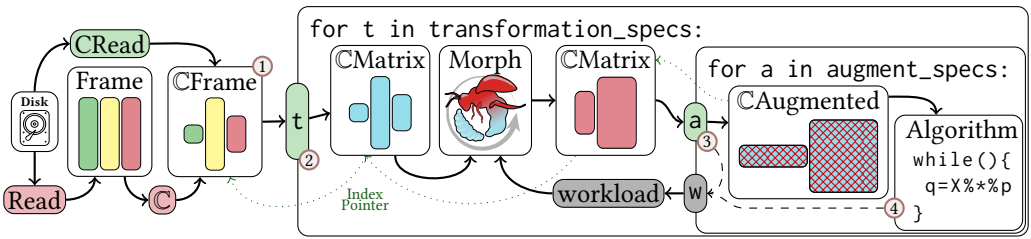


\begin{tikzpicture}[ampersand replacement=\&]


\begin{scope}[yshift = -3cm]
    \node[bbt1](IO2){\input{fig/prim/disk.tex}};
    \node[bbt1, xshift = .5cm*2](F2){\begin{tikzpicture}[ampersand replacement=\&]
    \node{
        \resizebox{1cm}{1cm}
        {
        \begin{tikzpicture}[ampersand replacement=\&]
            \matrix[Mtiles,
                column  1/.style={every node/.style={fulln, minimum width=1em, minimum height=4em, fill=green!50}},
                column  2/.style={every node/.style={fulln, minimum width=1em, minimum height=4em, fill=yellow!50}},
                column  3/.style={every node/.style={fulln, minimum width=1em, minimum height=4em, fill=red!50}},
             ](TILED){
                \  \& \  \& \ \\
            };
        \end{tikzpicture}
        }
    };
    \node at(0,0.7){Frame};
\end{tikzpicture}};
    \node[bbt2, fill = green!30](Cread)at ([yshift = 0.3cm]F2.north){CRead};
    \node[bbt2, fill = red!30](read)at ([yshift = -0.3cm, xshift = -1cm]F2.south){Read};

    \draw[->, thick](IO2) to[out = 90, in = 180] (Cread);
    \draw[->, thick](IO2) to[out = -90, in = 90] (read);
    \draw[->, thick](read) to[out = 0, in = -90] ([xshift = -0.2cm]F2.south);

    \node[bbt1, xshift = 1.2cm*2](CF2){\begin{tikzpicture}[ampersand replacement=\&]
    \node{
        \resizebox{1cm}{1cm}
        {
        \begin{tikzpicture}[ampersand replacement=\&]
            \matrix[Mtiles,
                column  1/.style={every node/.style={fulln, minimum width=1em, minimum height=1em, fill=green!50}},
                column  2/.style={every node/.style={fulln, minimum width=1em, minimum height=4em, fill=yellow!50}},
                column  3/.style={every node/.style={fulln, minimum width=1em, minimum height=2em, fill=red!50}},
             ](TILED){
                \  \& \  \& \ \\
            };
        \end{tikzpicture}
        }
    };
    \node at(0,0.7){$\mathbb{C}$Frame};
\end{tikzpicture}};
    
    \draw[->, thick](Cread) to[out = 0, in = 90] (CF2);

    \node[bbt2, fill = red!30](C)at ([yshift = -0.3cm, xshift = -0.7cm]CF2.south){$\mathbb{C}$};

    \draw[->, thick]([xshift = 0.2cm]F2.south) to[out = -90 , in = 180] (C);
    \draw[->,thick](C) to[out = 0, in =-90] ([xshift = -0.2cm]CF2.south);

    \begin{scope}[yshift = 0.28cm, xshift = -0.4cm]
        
        \node[bbt1, xshift = 2.5cm*2](M2){
\begin{tikzpicture}[ampersand replacement=\&]
    \node{
        \resizebox{1cm}{1cm}
        {
            \begin{tikzpicture}[ampersand replacement=\&]
                \matrix[Mtiles,
                column  1/.style={every node/.style={fulln, minimum width=1em, minimum height=1em, fill=blue!50}},
                column  2/.style={every node/.style={fulln, minimum width=1em, minimum height=4em, fill=blue!50}},
                column  3/.style={every node/.style={fulln, minimum width=1em, minimum height=2em, fill=blue!50}},
             ](TILED){
                \  \& \  \& \ \\
            };
            \end{tikzpicture}
        }
    };
    \node at(0,0.7){$\mathbb{C}$Matrix};
\end{tikzpicture}
};

        \node[bbt1, xshift = 3.2cm*2](Mor){\input{fig/prim/morph.tex}};
        
        \draw[->, thick] ([xshift = 0.2cm] M2.south) to[out = -90, in = -90]([xshift = -0.2cm] Mor.south);
        \node[bbt1, xshift = 4.cm*2](M3){\begin{tikzpicture}[ampersand replacement=\&]
    \node{
        \resizebox{1cm}{1cm}
        {
            \begin{tikzpicture}[ampersand replacement=\&]
                \matrix[tiles,
                column  1/.style={every node/.style={fulln, minimum width=1em, minimum height=1.8em, fill=red!50}},
                column  2/.style={every node/.style={fulln, minimum width=2em, minimum height=3em, fill=red!50}},
             ](TILED){
                \  \& \   \\
            };
            \end{tikzpicture}
        }
    };
    \node at(0,0.7){$\mathbb{C}$Matrix};
\end{tikzpicture}};

        \draw[->, thick](Mor) to[out = 0, in = 180](M3);

        \begin{scope}[xshift = 0.7cm, yshift = -0.1cm]
            
            \node[bbt1, xshift = 5cm*2, yshift = -1em- 0.1cm](A2){\begin{tikzpicture}[ampersand replacement=\&]
\newcommand{\augNode}[1]{
    \node[fulln, minimum width=4em, minimum height=#1em, fill=blue!50](N){ };
    \node[fulln, minimum width=4em, minimum height=#1em, fill=blue!50, pattern=crosshatch, pattern color=red,](N){ };  
}
        \node{
            
            \resizebox{2cm}{1cm}
            {
                \begin{tikzpicture}[ampersand replacement=\&]
                    \matrix[Mtiles,
                ](TILED){
                    \augNode{1.2}  \& \augNode{4} \\
                };
                \end{tikzpicture}
            }
        };
        \node at(0,0.7){$\mathbb{C}$Augmented};
    \end{tikzpicture}};
            \node[bbt1, xshift = 6cm*2, yshift = -1em- 0.5cm](Al2){\input{fig/prim/algo.tex}};
            
            \draw[->, thick]([yshift = 0.6cm]A2.east) to[out = 0, in = 90] ([xshift=-0.3cm]Al2.north);

            \node[ anchor = west](L1) at ([yshift = 0.2cm, xshift = -0.4cm]M2.north west){\texttt{for t in transformation\_specs:}};
            \node[ anchor = west](L2) at ([yshift = 0.2cm, xshift = -0.4cm]A2.north west){\texttt{for a in augment\_specs:}};
        
            \begin{scope}[on background layer]
                \node[bbt1, fit = (L2) (A2) (Al2)](BBL2){}; 
                \node[bbt1, fit = (L1) (L2) (M3) (M2) (Mor) (A2) (Al2) (BBL2)](BBL1){};
                \node[bbt1, fit = (L2) (A2)  (Al2)](BBL2){};
        
                \node[bbt1, minimum height = 1cm, minimum width = 0.3cm, fill = green!30](TLoopAccess)at (BBL1.west){\texttt{t}};
            
                \node[bbt1, minimum height = 0.5cm, minimum width = 0.3cm, fill = green!30](AugmentAccess)at ([yshift = 0.24cm]BBL2.west){\texttt{a}};
                \node[bbt1, minimum height = 0.5cm, minimum width = 0.3cm, fill = black!30](WorkloadAccess)at ([yshift = -0.7cm]BBL2.west){\texttt{w}};
            
            \end{scope}
        \end{scope}
    \end{scope}

    \draw[->, thick](M3) to[out = 0, in = 180] (AugmentAccess);
    \draw[->, thick](AugmentAccess) to[out = 0, in = 180] ([yshift = 0.26cm]A2.west);
    \draw[->, thick](CF2) to[out = 0, in = 180] (TLoopAccess);
    \node[bbt2, fill = black!30](WL)at([xshift = -1.cm]WorkloadAccess.west){workload};

    \draw[-> , thick](WorkloadAccess) to[out = 180, in = 0](WL);
    \draw[-> , thick](WL) to[out = 180, in = -90]([xshift=0.2cm]Mor.south);

    \draw[->, thick](TLoopAccess) to (M2);
\end{scope}

\node[roundnodeb]at(CF2.north east){1};
\node[roundnodeb]at(TLoopAccess.south east){2};
\node[roundnodeb](Three)at(AugmentAccess.south east){3};
\node[roundnodeb](Four)at([yshift = 0.3cm]Al2.south west){4};

\draw[->, dashed](Three) to[out = -45, in= 0] (WorkloadAccess.east);
\draw[->, dashed](Four) to[out = -180, in= 0] ([xshift = -2cm]Four) to[out = -180, in= 0] (WorkloadAccess.east);

\draw[->, dotted, color = green!60!black]([xshift = -0.2cm]M2.south) to[out = -110, in = -70] ([xshift = 0.2cm]CF2.south);
\draw[->, dotted, color = green!60!black]([xshift = -0.4cm]M3.south) to[out = -130, in = -50] ([xshift = -0.2cm]M2.south);
\draw[->, dotted, color = green!60!black]([yshift = -0.2cm]A2.north west) to[out = -180, in = -0] ([yshift = -0.2cm]M3.north east);

\begin{scope}
    \node[text = green!60!black] at([xshift = -0.2cm, yshift = -0.5cm]M2.south){\tiny Index};
    \node[text = green!60!black] at([xshift = -0.2cm, yshift = -0.65cm]M2.south){\tiny Pointer};
\end{scope}

\end{tikzpicture}

\caption{BWARE: Data-Centric ML Pipeline with Compiler Introduced Compression}
\label{fig:MLCompressedPipeline}
\end{figure}

\textbf{Example:}{\textit{ \cref{fig:MLCompressedPipeline} shows the compiled execution plan for the script from \cref{fig:MLPipeline}. In stage }\nodeLabelb{1}\textit{ the compiler injects frame compression, depending on the input file, it either compresses an uncompressed input frame or directly reads a $\mathbb{C}$frame. The $\mathbb{C}$frame is transformed into a matrix, where index structures can be reused, and most transformation costs are $\mathcal{O}(1)$. For stage }\nodeLabelb{2}\textit{ the compiler introduces a morphing instruction to optimize the compressed format according to the workload extracted from the linear algebra program and used operations in stages }\nodeLabelb{3} \textit{\&} \nodeLabelb{4}\textit{ (we use DSL-based primitives for augmentation and algorithms). For some operations, the optimizer also introduces morphing into the algorithms in }\nodeLabelb{4}\textit{.}}

\setlength{\columnsep}{0.4cm}%

\section{Experiments}
\label{sec:experiments}

Our experiments study various properties of workload-aware compression. We start with the sizes and compression speed of frames.
Then, we compare lossless and lossy approaches to feature transformations. 
We show how our solution scales evaluating polynomial feature engineering performance and highlight a word embedding NLP example with a fully connected layer.
Further experiments show end-to-end ML algorithms using both lossless and lossy feature engineering. 
We also evaluate a data-centric ML pipeline that iterates through multiple feature transformations. 
Finally, we compare the transformation performance to other systems.

\subsection{Experimental Setting}

\textbf{HW/SW Setup:} All experiments are conducted on a server with two Intel Xeon Gold 6338 \@2.0-3.2\,GHz (64 cores, 128 threads) and 1\,TB 3200\,MHz DDR4 RAM (peak performance is 6.55\,TFLOP/s), $16\times$ SATA SSDs in RAID 0 for the datasets, and an Intel Optane SSD DC P5800X for the programs, scripts, and local evictions (in case live variables exceed the buffer pool size). Our software stack comprises Java 17.0.11, SystemDS, Hadoop 3.3.6, and Spark 3.5.0.

\begin{table}[!t] \setlength\tabcolsep{8.2pt}
    \caption{Used Datasets and ML Tasks.\label{tab:datasets}}
    \vspace{-0.4cm}
    \begin{tabular}{r|r r r r c }
        \toprule
        \textbf{Dataset Name}                               & \textbf{\# Rows}        & \textbf{\# Cols}   & \textbf{Categorical} & \textbf{Numeric} & \textbf{Task}\\
        \midrule
        Adult        \cite{Data-adult}              & \numprint{    32561} & \numprint{ 15}  &            9   &              6 & Binary\\
        CatInDat     \cite{Data-cat, Data-cat-ii}   & \numprint{   900000} & \numprint{ 24}  &           16   &              8 & Binary\\
        Criteo Day 0 \cite{Data-criteo}             & \numprint{195841983} & \numprint{ 39}  &           25   &             14 & Binary\\
        Crypto       \cite{Data-crypto}             & \numprint{ 24236806} & \numprint{ 10}  &            1   &              9 & Regression\\ 
        KDD98        \cite{Data-kdd}                & \numprint{    96367} & \numprint{481}  &          135   &            334 & Regression\\ 
        Santander    \cite{Data-santander}          & \numprint{   200000} & \numprint{201}  &            0   &            201 & Binary\\ 
        HomeCredit   \cite{Data-HomeCredit}         & \numprint{   307511} & \numprint{121}  &           16   &            105 & Binary\\
        Salaries     \cite{Data-salaries}           & \numprint{      397} & \numprint{  6}  &            3   &              3 & Binary\\ 
        \midrule
        AMiner V16 \cite{TangZYLZS08-DBLPAbstracts} & $\approx$\numprint{4000000} & \numprint{1000} & \numprint{1000} & 0 & Word Embed\\
        \bottomrule
    \end{tabular}
\end{table}

\textbf{Baselines:} As baselines, we utilize SystemDS uncompressed I/O and operations (ULA), and compare to AWARE~\cite{BaunsgaardB23} for lossless matrix compression, as well as ML systems (TensorFlow 2.15~\cite{tensorflow}, SKLearn 1.4.1~\cite{scikit-learn, sklearn_api}), data management systems (Pandas 2.1~\cite{Pandas}, Polars 0.20~\cite{POLARS}), and generic compression systems (ZStd 1.5.5-4 ~\cite{zstd}, Snappy 1.1.10.3~\cite{Snappy}).  

\subsection{Datasets}
We use multiple datasets, each having different data types and transformation requirements. Adult~\cite{Data-adult} (also called Census) is a sample from the person records in a 1990 U.S. census. CatInDat~\cite{Data-cat, Data-cat-ii} (Cat) is a synthetic dataset from Kaggle that contains categorical features for predicting cat ownership, we combined the two competition datasets. Criteo~\cite{Data-criteo} is a dataset of millions of display advertisements for predicting which ads were clicked. Criteo10M is the first 10 million rows from Criteo. Crypto~\cite{Data-crypto} is a Google competition dataset for time series forecasting of crypto-currencies. KDD98~\cite{Data-kdd} is a knowledge discovery competition dataset from 1998. Salaries~\cite{Data-salaries} is a small dataset containing the salaries of professors in a U.S. college. Santander~\cite{Data-santander} is another Kaggle competition to predict customer transactions. HomeCredit~\cite{Data-HomeCredit} predicts how likely each applicant is to repay a loan. AMiner V16~\cite{TangZYLZS08-DBLPAbstracts} contains $\approx$4 million abstracts from various conferences, we preprocessed AMiner by removing non-English abstracts, equations, and symbols. The datasets are summarized in \cref{tab:datasets}. Categorical and numeric is the number of respective feature types. The tasks are split into regression, binary classification, and word embedding tasks.

\subsection{Frame Compression and I/O}

\begin{figure}[!t]
    \includegraphics[scale=0.99]{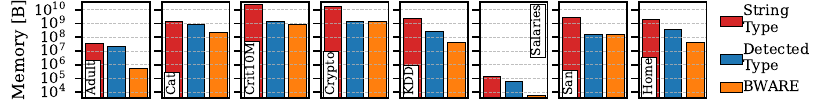}

    \includegraphics[scale=0.99]{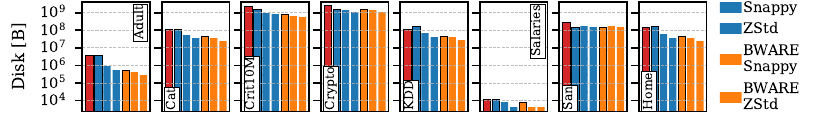}

    \includegraphics[scale=0.99]{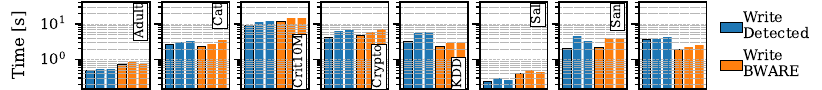}

    \includegraphics[scale=0.99]{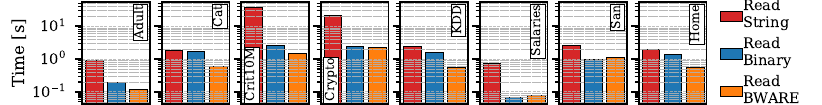}
    
    \vspace{-0.25cm}
    \caption{\label{fig:CompressionSizeFrame}Frame Compression.}
    \Description{None}
\end{figure}

We first evaluate the sizes of compressed frames and I/O performance in \cref{fig:CompressionSizeFrame}.

\textbf{In-Memory Size:} \cref{fig:CompressionSizeFrame}'s top row shows three different measures of the in-memory compressed frame. First, \emph{String} represents a frame with the default generic string values without exploiting the values types of the columns. Second, \emph{Detect} automatically detects the value types. \emph{Detect} achieves in-memory size reductions from 1.5x to 18x across the datasets compared to \emph{String}. However, BWARE's compressed frame improves it by 19x to 65x. Comparing BWARE with \emph{Detect} shows additional improvements of 1.09x to 43x. A low ratio relative to \emph{Detect} occurs in cases with continuous values and high cardinality, such as Salaries or Crypto. The results show that BWARE can keep larger frames using less memory and guarantees (except for boolean data) less than or equal sizes to \emph{Detect}. Interestingly, BWARE reduces allocation even in the tiny Salaries dataset.

\textbf{On-Disk Size:} The second row in \cref{fig:CompressionSizeFrame} shows the size of the datasets on disk. The first column in each figure shows the original allocations (CSV files). The second column contains our serialized detected frame saved in HDFS sequence files, with tiles of 16K rows. We see that there can be an overhead in storing the tiles. The worst case is KDD98 going from 112MB CSV to 171MB, a 1.5x increase. If HDFS's block-wise compression is enabled, then using Snappy improves KDD98's binary files to 63MB, while ZStd is better with 36MB. Enabling our compressed writers in BWARE, we get 45MB, and ZStd's nested compression 24MB an on-disk reduction of $4.6$x. The general conclusion is that BWARE performs almost equally to other compression frameworks for individual block compression, and we can recursively stack compression schemes for improvements.

\textbf{Write/Read Time:} The third row in \cref{fig:CompressionSizeFrame} shows the execution time to write the different formats. The writing time includes schema detection, schema application, and compression if applicable. We see a moderate overhead for compression but it is comparable to other compressors. The last row in \cref{fig:CompressionSizeFrame} shows the reading performance from CSV files as well as uncompressed and compressed binary files. We see that reading text formats should be avoided, but sometimes it is competitive with our binary format (e.g., in Cat). The BWARE reader performs similarly or better than the uncompressed binary reader, even in incompressible cases like Crypto and Santander. The exception is the tiny Salaries dataset, where the binary reader is the fastest.

\subsection{Compressed Feature Transformations}

\begin{wrapfigure}{r}{8.5cm}
    \vspace{-0.3cm}
    \includegraphics[scale=0.99]{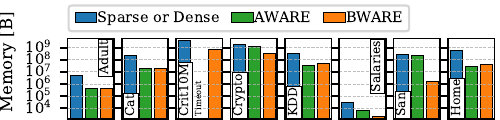}

    \includegraphics[scale=0.99]{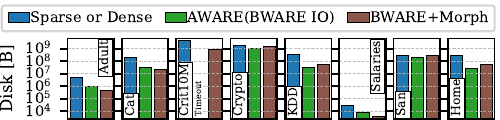}

    \includegraphics[scale=0.99]{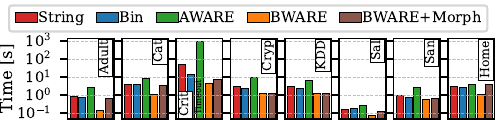}

    \vspace{-0.25cm}
    \caption{\label{fig:CompressedTransformEncodeLossless}Lossless Transform-Encode.}
    \Description{None}
    \vspace{-0.2cm}
\end{wrapfigure}
Next, we evaluate feature transformations. We start with lossless followed by lossy encodings.

\textbf{Lossless Encoding:} We one-hot-encode all categorical, and pass-through numeric features in a lossless encoding. With this scheme, we copy over values of Crypto and Salaries because all values are numeric, while most columns in Cat and Criteo are one-hot encoded. \cref{fig:CompressedTransformEncodeLossless} shows the performance. The rows include (1) the matrix's last size in memory, (2) the saved size on disk, and (3) the execution time of \texttt{transformencode} plus compression or morphing. 
AWARE and BWARE use less memory than a default sparse or dense matrix, even in the incompressible Crypto and Santander. AWARE's allocation is close but consistently smaller than BWARE. However, BWARE reuses the index structure from the compressed frame arrays in cases with a 1-to-1 mapping from the frame column's values. We further see that BWARE's on-disk representation is slightly worse than AWARE (which we extended to use BWARE's I/O operations). 

\textbf{Lossless Time:}
\emph{String} encodes string types as input, while Binary (F-M) encodes using the detected types. AWARE (F-M-$\mathbb{C}$M) encodes detected types followed by compression from scratch, and BWARE (F-$\mathbb{C}$M) uses compressed transform encoding. BWARE is faster than the other solutions, with exceptions in the incompressible cases of Crypto and Santander. AWARE's compression of Criteo 
\begin{wrapfigure}{r}{8.5cm}
    \vspace{-0.1cm}
    \includegraphics[scale=0.99]{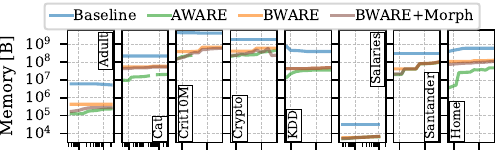}

    \includegraphics[scale=0.99]{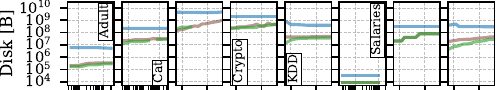}

    \includegraphics[scale=0.99]{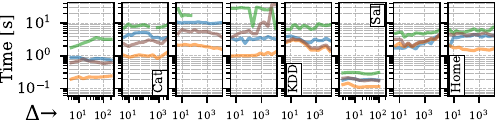}

    \vspace{-0.25cm}
    \caption{\label{fig:CompressedTransformEncodeLossy}Lossy Transform-Encode Size and Time.}
    \Description{None}
	\vspace{-0.257cm}
\end{wrapfigure}
shows what happens when the rediscovering of column correlations dominates. In essence, Criteo is encoded into millions of perfectly correlated columns because of the one-hot encoding. AWARE tries to discover the correlation and starts co-coding. However, due to millions of columns with perfect correlation and each co-coding candidate takes time to verify, we timeout the runs at 1000 seconds.

\textbf{Lossy Encoding:} \cref{fig:CompressedTransformEncodeLossy} shows results with different $\dv$ (\#bins) for numeric ranges or categorical hash buckets on the x-axis. BWARE without morphing uses the DDC compression of the compressed \texttt{transformencode}, while BWARE with morphing additionally morphs the compression scheme after the transformation.
AWARE, BWARE, and BWARE+Morphing use less space than the uncompressed baseline. AWARE returns better-compressed results than BWARE because it has a larger exploration space, while BWARE is more optimized for speed and reuse. When writing to disk, we always use morphing to improve the allocation. We observe that the morphed allocation is close to AWARE's saved format. The results show that BWARE is faster across all datasets while being on-par for Santander. BWARE yields improvements of 2x to 20x over AWARE and 1x to 5x over the baseline SystemDS.

\begin{wrapfigure}{r}{0.4\linewidth}
        \vspace{-0.3cm}
        \includegraphics[scale=0.99]{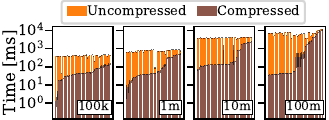}
        
        \vspace{-0.3cm}
        \caption{\label{fig:compressedInputFrameOrNot} Compressed Input Frames.}
        \Description{None}
        \vspace{-0.2cm}
\end{wrapfigure}
\textbf{Encoding Time Breakdown:} All experiments, so far, used uncompressed frame inputs. If the input frame's columns are compressed, the asymptotic runtime changes to constant for some transformations. 

\cref{fig:compressedInputFrameOrNot} shows the parallel lossless transform encode time of individual columns of Criteo for different numbers of rows with columns sorted by compressed execution time. Some columns in Criteo are compressed, while others have many distinct values and, therefore, are uncompressed. The constant encoding time can be seen in the plateau of the first 50\% of the columns in the plot. The constant groups take $\sim40$ms, except for 10m where it consistently is ~$\sim100ms$.
The following 25\% have to change their compressed index structures, and the final 25\% are uncompressed. The two fast columns in uncompressed are boolean columns. Since our hardware setup has a high degree of parallelism, the total encoding time is equal to the tallest bar, while a single-threaded execution is equal to the integral of the colored areas. Therefore, the end-to-end difference between F-$\mathbb{C}$M and $\mathbb{C}$F-$\mathbb{C}$M is small if $\mathbb{C}$F contains incompressible columns.

\subsection{Compressed Word Embeddings}

\begin{wrapfigure}{r}{3in}
    \vspace{-0.25cm}
    \includegraphics[scale=0.99]{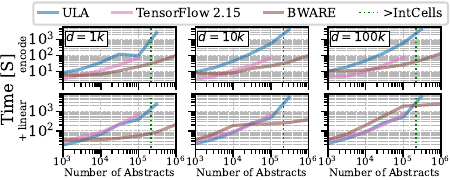}
    \caption{\label{fig:woredEmbResults}Compressed Word Embedding}
    \Description{None}
\end{wrapfigure}

\cref{fig:woredEmbResults} shows the performance of encoding already tokenized abstracts from DBLP~\cite{TangZYLZS08-DBLPAbstracts}, capped at a maximum of \numprint{1000} tokens. All plots show the total execution time of 10 repetitions of performing word embedding with word2vec \cite{MikolowGBPJ18} embeddings trained on Wikipedia. The first row represents word embedding only, while the second row adds a fully-connected neural network layer with ReLu activation on the embedded outputs. The columns show increasing numbers of unique tokens ($d$) allowed, starting at $d = 1k$ and increasing to $100k$. The x-axis on all plots is the number of abstracts encoded, while the y-axis is execution time. We observe that ULA is slower at embedding but as fast as TensorFlow once the neural network layer is added. ULA catches up because of efficient sparse linear algebra not leveraged in TensorFlow. TensorFlow and ULA are not affected by increasing $d$, while BWARE is. BWARE shows the best performance in all cases in embedding time and scales further than the other implementations. When adding the neural network layer, the performance is slower in cases where the number of abstracts is lower than $d$. However, once the number of abstracts is larger than $d$, BWARE asymptotically and empirically outperforms all the other implementations.

\subsection{ML Algorithm Performance}

To quantify BWARE's effect, we evaluate a conjugate gradient linear model training with different lossless, lossy, and feature engineering pipelines showcasing individual effects BWARE has on the combined performance of end-to-end ML pipelines.

\begin{wrapfigure}{r}{8.4cm}
    \centering
    \includegraphics[scale=0.99]{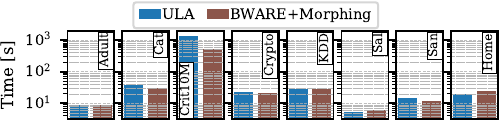}
    \caption{\label{fig:baslineLM}LM Conjugate Gradient Baseline.}
    \Description{None}
\end{wrapfigure}
\textbf{Lossless:} \cref{fig:baslineLM} shows the performance of training a linear regressorconjugate gradient method with L2 regularization. The max iterations is set to min(\#col, 1000). The algorithms are sparse-safe, allowing the baseline to take full advantage of sparse linear algebra. We observe a small slowdown using BWARE in some of the cases of 22\% in KDD and 25\% in Home. However, Criteo (with 10 million rows) improves by 2x from \numprint{1368} to \numprint{681} seconds. Incompressible cases are expected to have a small overhead in analysis and compression, but we observe that both the incompressible cases of Crypto and Santander keep the same execution time because the compressed transformations fall back to uncompressed representations. We do not show the accuracy of models because the results of both solutions are the same. However, as an example, the method scores 78.9 AUC for Cat on Kaggle with such a simple model, while the top score is 80\%~\cite{Data-cat}.

\begin{wrapfigure}{r}{3in}
    \includegraphics[scale=0.99]{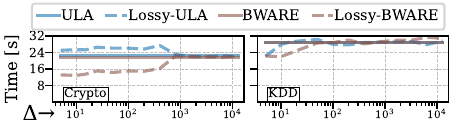}    
    \caption{\label{fig:LMLossy}LM Conjugate Gradient Loss w/ Increasing $\dv$.}
    \Description{None}
\end{wrapfigure}

\textbf{Lossy:} \cref{fig:LMLossy} shows the results when controlling the number of distinct values through lossy transformations. The Crypto dataset is almost purely numeric and a dense dataset. When varying $\dv$, we observe baseline performance similar to the lossless solution. The performance is expected since there is no benefit from reducing the number of unique values in uncompressed linear algebra. BWARE is able to exploit the reduced number of unique values, with a slight increase in run times when $\dv$ increases. There are cases that do not benefit, such as KDD, where only extreme values of $\dv$ yield performance improvements. Lower $\dv$ generally makes models fit worse, but not always, and sometimes lower $\dv$ can have a positive regularization effect that gives better accuracies. For KDD, the break-even point of lossy and lossless accuracies is $\dv=800$. The model is able to fit just as well, and sometimes better, on some lossy inputs using only quantization. The results indicate that $\dv$ has a positive impact on runtime with an unknown negative or positive impact on accuracy. Hence, one should perform automated feature engineering.

\begin{wrapfigure}{r}{0.34\linewidth}
    \vspace{-0.2cm}
    \includegraphics[scale=0.99]{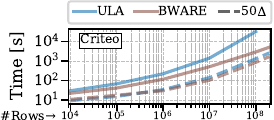}
    
    \vspace{-0.2cm}
    \caption{\label{fig:LMScaling}LM Training Scaling Rows.}
    \Description{None}
	\vspace{-0.2cm}
\end{wrapfigure}
\textbf{Scaling:} \cref{fig:LMScaling} shows the scalability of BWARE in terms of performance on larger subsets of the Criteo dataset. We observe a starting 2x performance improvement in \cref{fig:baslineLM} at $10^4$ rows. The improvement increases in all cases until $10^9$ rows. BWARE is a substantial 11x faster at $10^8$ rows, improving from \numprint{31792} to \numprint{2880} sec. With lossy encodings of Criteo, both ULA and BWARE show similar improved performance, with an increasing gap for more rows.

\begin{wrapfigure}{r}{0.5\linewidth}
    \vspace{-0.2cm}
    \includegraphics[scale=0.99]{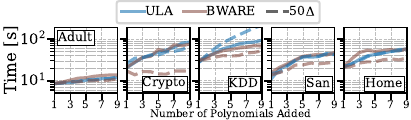}
    \vspace{-0.2cm}
    \caption{\label{fig:poly}Kernel Trick Polynomial Regression.}
    \Description{None}
    \vspace{-0.2cm}
\end{wrapfigure}
\textbf{Polynomial Regression:} \cref{fig:poly} shows the polynomial regression (leveraging the kernel trick) results. These results indicate that BWARE facilitates polynomial feature engineering with very moderate costs and sometimes improvements. The best case is Crypto, where the polynomial features do not affect the execution time when combined with lossy feature transformations. The other datasets do not have significant performance improvements. BWARE performs poorly with lossless feature transformations in the Home dataset at low polynomials because the dataset contains a few incompressible columns (see \cref{fig:RelativeDistinct}), which do not amortize with lossless feature transformations. However, once these columns are transformed in a lossy manner, the performance is good.

\begin{wrapfigure}{r}{2.11in}
    \vspace{-0.3cm}
    \includegraphics[scale=0.99]{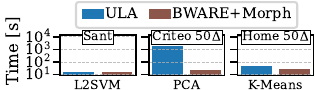}
    \vspace{-0.2cm}
    \caption{\label{fig:OtherAlgorithms}Various ML Algorithms.}
    \Description{None}
    \vspace{-0.3cm}
\end{wrapfigure}

\textbf{Other Algorithms:} AWARE already studied the impact of compression on multiple linear-algebra-based ML algorithms, which we inherit for BWARE. \cref{fig:OtherAlgorithms} shows the performance of a few other algorithms. First, BWARE shows equal performance to ULA for L2SVM on Santander. PCA on Criteo with a lossy transformation shows an 83x improvement in execution time. This relative improvement in performance can be arbitrarily large depending on $\dv$ because PCA is asymptotically faster in compressed space. Finally, BWARE shows a solid 2x improvement for K-means on Home using a lossy transformation.

\begin{table}[!h] \setlength\tabcolsep{13.pt}
    \centering
    \caption{\label{tab:pipeline}Pipeline LM: 8 Transform Encode \& 8 Polynomials.}
	\vspace{-0.4cm}
    \begin{tabular}{r |r|c|c|c}
        \toprule
        \textbf{Dataset} & \textbf{Measure}        & \textbf{ULA} & \textbf{AWARE}  & \textbf{BWARE} \\
          \midrule
          \textbf{KDD98} & Execution Time           & \numprint{  654}s & \numprint{  452}s  & \numprint{251}s\\
           & Instructions   & \numprint{  110}{\tiny$\cdot10^{12}$} & \numprint{  100}{\tiny$\cdot10^{12}$}  & \numprint{ 44}{\tiny$\cdot10^{12}$}\\
        &  Instructions per Cycle  & \numprint{ 0.94}  & \numprint{ 2.59}   & \numprint{2.68}\\
        &  L1-dcache-miss & \numprint{ 7792}{\tiny$\cdot10^9$} & \numprint{1740}{\tiny$\cdot10^9$}  & \numprint{786}{\tiny$\cdot10^9$} \\ 
        &  Compress/Morph        &  --- & \numprint{148}s & \numprint{21.9}s \\ 
        &  Transform-Encode &  \numprint{74.1}s & \numprint{59.9}s & \numprint{7.95}s \\
        &  Energy Consumption~\cite{Noureddine22} & \numprint{338}kJ & \numprint{185}kJ & \numprint{92}kJ \\
        \midrule
        \textbf{Home} &  Execution Time & \numprint{431}s & \numprint{266}s & \numprint{160}s\\
        \textbf{Adult} & Execution Time &  \numprint{19.9}s & \numprint{22.1}s & \numprint{16.6}s\\
        \textbf{Santander} & Execution Time &  \numprint{489}s & \numprint{376}s & \numprint{374}s\\
        \textbf{Cat} & Execution Time &  \numprint{467}s & \numprint{170}s  & \numprint{63}s\\
        \bottomrule
   \end{tabular}
	\vspace{.3cm}
\end{table}

\subsection{Data-centric ML Pipeline}

\cref{tab:pipeline} shows the execution time and characteristics of a full, end-to-end, data-centric ML pipeline similar to \cref{fig:MLCompressedPipeline}. The table contains results from a pipeline performing a grid search of hyper-parameters with eight different $\dv$ ranging from 5 to 480 and eight polynomials from 1 to 8. The pipeline uses two outer loops: the first performing \texttt{transformencode}, and the second polynomial feature construction. The top half of the table contains performance numbers for the KDD dataset. AWARE is 1.45x faster than ULA, and BWARE further improves by 1.8x. BWARE is the fastest because it reuses intermediate compressed representations through the feature transformations, and AWARE redundantly rediscovers correlated columns. BWARE's handling of polynomial features is also used by AWARE in this experiment. AWARE and BWARE compression also show better cache locality than ULA, which decreases L1 cache misses by an order of magnitude, and in turn increases instructions performed per CPU cycle. Furthermore, BWARE improves energy consumption due to more efficient data types and reduced cache misses, because data access is a major energy consumer~\cite{Horowitz14}. The bottom half of the table shows the results of the same pipeline on Home, Adult, Santander, and Cat. BWARE is the fastest in all cases. AWARE also does well, except a small overhead on Adult where the compression overhead cannot be amortized.

\subsection{Comparisons with Other Systems}
\begin{wrapfigure}{r}{7.7cm}
    \centering
    \vspace{-1cm}
    \includegraphics[scale=0.99]{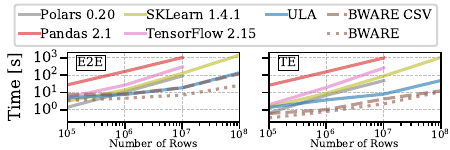}
    
    \vspace{-0.3cm}
    \caption{\label{fig:RelatedSystems}Performance Comparison with Other Systems.}
    \Description{None}
    \vspace{-0.25cm}
\end{wrapfigure}
Finally, \cref{fig:RelatedSystems} compares the performance of \texttt{transformencode} on Criteo with other systems. The left plot (E2E) is end-to-end times with CSV parsing, and right (TE) is only \texttt{transformencode}. ULA and BWARE have high startup times in E2E, and JIT compilation benefits in Java are limited for small inputs. However, for larger sizes, ULA and BWARE outperform the other systems. Polars is the only system with a dense matrix result but with \texttt{UINT8} encoded columns. All other systems use sparse transformations to run until $10^7$ rows. At $10^7$ rows, BWARE is 4.7x (E2E) and 11.4x (TE) faster than Polars. BWARE's E2E times are equal to ULA when reading CSV, which is still good because it yields a compressed output for subsequent operations (F-$\mathbb{C}$M). We also included a dotted line for BWARE reading a compressed frame from disk for compressed encoding ($\mathbb{C}$F-$\mathbb{C}$M). For more than $2^{32}$ cells (max integer), many of the systems crash. SK-learn can scale further, but BWARE is 11.9x (E2E) and 76.2x (TE) faster than SK-learn at $10^8$ rows.


\balance
\section{Related Work}
\label{sec:relatedWork}

BWARE is connected to multiple related fields, including database compression, matrix compression, and specific techniques for workload-awareness and data reorganization.

\textbf{Database Frame Compression:} Compressing heterogeneous data is common in the data management literature. Tabular data in databases is compressed with schemes that exploit entire columns having the same datatype~\cite{AbadiBH13,StonebreakerABCCFLLMOORTZ05}. Systems often employ variations of five lightweight encodings techniques~\cite{DammeDJW17, DammeUHJDL19}: Frame-of-reference (FOR)~\cite{ZukowskiHNB06}, delta encoding (DELTA)~\cite{ZukowskiHNB06}, dictionary encoding (DDC)~\cite{MullerCF14, WestmannKHM2000, AntoshenkovLM96}, run-length encoding (RLE)~\cite{AbadiMF06} and null suppression (NS)~\cite{LemireB15,WestmannKHM2000}. General-purpose, heavy-weight compressors are also applied to compress any modality of data. Examples include Snappy~\cite{Snappy} and Zstd~\cite{zstd}. Most systems support storing data in compressed formats that can combine multiple techniques. BtrBlocks~\cite{KuschweskiSAL23} is a recent work showcasing the effectiveness of nesting compression techniques with highly efficient SIMD decompression. Other examples include Parquet~\cite{Apache-Parquet}, HDF5~\cite{hdf5}, and SciDB~\cite{StonebreakerBPR11} for storage, as well as Arrow~\cite{Apache-Arrow, Apache-Arrow-dictionary} for data transfer. There are many dedicated compression techniques for specific value types, for instance, for strings in Pattern-Based Compression~(PBC)~\cite{ZhangSYMXJLSZL23} that decomposes string values via entropy encoding (Huffman coding~\cite{HuffmanCoding1952,RamanS06}), and FSST~\cite{BonczNL20}. These fine-grained methods, such as white box compression \cite{GhitaGB20}, share some commonalities but are orthogonal. We do similarly read and write compressed data blocks~\cite{LangMFBN} in BWARE, but we differentiate by enabling algorithms to directly process the heterogeneous compressed formats without decompression.

\textbf{Lossless Matrix Compression:} Compression of numeric data has a long history as well. The most studied type of compression is integer-based compression~\cite{KuschweskiSAL23, DammeDJW17, DammeUHJDL19, LemireB15} while floating points with exponent and mantissa pose some difficulties. XOR~\cite{RatanaworabhanJB06} compression used in Gorilla~\cite{PelkonenFTCHMV15} is a technique specific for floating point data, followed by advancements in Chimp~\cite{LiakosPK22} and even more recently in ALP~\cite{AfroozehKB23}. Sparsity-exploiting compression had full software package support already in 1990 with SparseKit~\cite{Saad90} using specialized data structures to exploit non-zero values (CSR, CSC, and COO). Sparsity exploitation is now commonly supported in most linear algebra frameworks such as IntelMKL (now behind OneAPI)~\cite{IntelMKL} and CuSparse~\cite{CuSparse}, but is still in active research~\cite{SommerBERH19, KhamisNNOS20, WilkinsonCD23}. UniSparse~\cite{LiuZDBRZ24} is a recent example of an MLIR-based~\cite{LattnerABCDPRSVZ20} sparse tensor system with a compiler optimizing and selecting various sparse formats by---similar to BWARE---decoupling the logical representations from the given user programs. \citeauthor{ZhangHK24} explores generating code that---similar to us injecting morphing---automatically converts intermediates to sparse tensors~\cite{ZhangHK24}. WACO~\cite{WonMEA23} is another recent sparsity-exploiting framework, that optimizes the sparse formats based on linear algebra and data inputs. Exploiting other types of redundancy for faster processing via compression is done in Tuple-oriented Compression (TOC)~\cite{FenganLYANPM19}, Compressed Linear Algebra (CLA)~\cite{ElgoharyBHRR18,ElgoharyBHRR16}, and AWARE~\cite{BaunsgaardB23}. TOC, CLA, and AWARE further took inspiration from sparse matrix compression~\cite{KourtisGK08, KarakasisGKGK13}. In contrast to existing work, we push compression through entire data-centric ML pipelines including feature transformations without recompressing data.

\textbf{Lossy Matrix Compression:} Mainstream ML systems rely mostly on homogeneous lossy compression partially because it retains regular dense data access. The lossy exploitation is done in two different directions. First, a common choice is uniformly encoding and processing all floating point values in reduced precision~\cite{IEEE_FLOAT_STANDARD}. Specialized custom data types are also explored with success such as Google's bFloat16~\cite{GOOGLE_BFLOAT16, kalamkarDNDKSDNJHJJAESAMBP2019}, Intel's Flexpoint~\cite{KosterWWNBCEHHK17} and NVIDIA's TF32\cite{A100}. Some solutions use multiple precisions for different operators~\cite{WangCBCG18, ZhangLKADLZ17}, even going as far as 1 bit aggregations in neural networks~\cite{SeideFDLY14, HubaraCSEB18}. Another group of lossy exploitation is dimensionality reductions via, for instance, auto encoders~\cite{IlkhechiCGMFSC20} or classic algorithms such as PCA~\cite{PearsonPCA1901} and t-SNE~\cite{HintonR02}. Sampling or coresets~\cite{AgarwalPV07} is another lossy approach allowing training with fewer mini batches~\cite{SuchRLSC20} or random samples for each batch~\cite{ParkQSM19}. Our approach exploits user-defined lossy feature transformations with lossless compression techniques to avoid trust concerns in result validity from lossy decisions.

\textbf{Workload-aware Compression:} The online workload-aware compression from AWARE~\cite{BaunsgaardB23}, tunes the compression based on workload characteristics of a linear algebra program. Others adapt similarly based on sparsity~\cite{WillowKA22}. Several systems also combine cost modeling of compressed size and query performance~\cite{VaradarajanBCDB14, BoissierJ19, CenKMK21, DammeUAJADL20, KimuraNS11}, but many of these techniques rely on an offline compression, unlike us, for selection and adapting to workloads~\cite{Boissier21}. Some related work even generates workload traces~\cite{PaulCKW22}. In contrast, BWARE does workload-aware compression through feature transformations in an online manner during runtime for performance tuning based on compile-time information.

\textbf{Data Reorganization:} Our morphing technique is primarily a data reorganization strategy. Prior work, like database cracking by \citeauthor{IdreosKM07_1} \cite{PirkPIMK14, HalimIKY12, IdreosMKG11, IdreosKM07_1, IdreosKM07_2}, also dynamically reorganizes data based on query workload. Other work dynamically chooses: (1) the physical design of storing data~\cite{ArulrajPM16}, (2) where to place tuples on distributed servers (e.g. Clay~\cite{SerafiniTEPAS16}), and (3) online deduplication of stored blocks~\cite{XuPSG17}. MorphStore~\cite{ DammeDJW17, HabichDUPKHL19} proposes a morphing wrapper enabling on-the-fly recompression of intermediate results with lightweight compression schemes for relational algebra. Unlike many related works, we perform workload-aware reorganization of matrices for data-centric ML pipelines.

\textbf{Compressed Operations:} There exist multiple other works for performing operations directly on compressed (matrix) formats. Factorized learning~\cite{KumarNP15, MaximilianOC16, KhamisNNOS20, Olteanu20} pushes ML workloads through joins, avoiding the materialization of denormalized tables. Grammar-based compressed operations \cite{FerraginaMGTKNST22, AbboudBBK23, RajatK24} also show good performance, specifically the CSRV representation~\cite{FerraginaMGTKNST22}. CLA~\cite{ElgoharyBHRR18, ElgoharyBHRR16} support multiple operations in compressed space, while AWARE~\cite{BaunsgaardB23} extended the operation support to full matrix multiplications. BWARE further extends the operations into feature transformations.
\section{Conclusions}
\label{sec:conclusions}

We introduced BWARE as a holistic, lossless compression framework for data-centric ML pipelines, which is fully integrated in SystemDS~\cite{BoehmADGIKLPR20,BoehmDEEMPRRSST16}. 
In this context, we push compression through feature transformations and feature engineering into the sources. We draw two main conclusions. First, this compression strategy is able to yield substantial runtime improvements because of repeated feature transformations and ML model training. Second, compressed feature transformations preserve information about structural redundancy, achieving improved compression ratios and thus, better data locality. Interesting future work includes support for more feature transformations (e.g., image data augmentation) and specialized, heterogeneous hardware accelerators.

\bibliographystyle{ACM-Reference-Format}
\bibliography{arXiv}

\end{document}